\pdfoutput=1
\documentclass[fleqn,10pt]{wlscirep}
\usepackage[utf8]{inputenc}
\usepackage[T1]{fontenc}
\usepackage[version=3]{mhchem}
\usepackage{xcolor}
\usepackage{tabularx}
\usepackage{rotating}
\usepackage{makecell}
\usepackage{caption}
\usepackage{subcaption}
\usepackage{amssymb}
\usepackage{hyperref}

\title{Establishing magneto-structural relationships in the solid solutions of the skyrmion hosting family of materials: GaV$_4$S$_{8-y}$Se$_{y}$}

\author[1, *]{Ale\v{s} \v{S}tefan\v{c}i\v{c}}
\author[1]{Samuel J. R. Holt}
\author[1]{Martin R. Lees}
\author[2]{Clemens Ritter}
\author[3]{Matthias J. Gutmann}
\author[4]{Tom Lancaster}
\author[1, +]{Geetha Balakrishnan}
\affil[1]{University of Warwick, Department of Physics, Coventry, CV4 7AL, United Kingdom}
\affil[2]{Institut Laue Langevin, 38042 Grenoble Cedex, France}
\affil[3]{ISIS Facility, Rutherford Appleton Laboratory, Oxfordshire, OX11 0QX, United Kingdom}
\affil[4]{Durham University, Department of Physics, South Road, Durham DH1 3LE, United Kingdom}

\affil[*]{Ales.Stefancic@psi.ch}
\affil[+]{G.Balakrishnan@warwick.ac.uk}


\begin{abstract}
    The GaV$_4$S$_{8-y}$Se$_y$ $(y = 0$ to $8)$ family of materials have been synthesized in both polycrystalline and single crystal form, and their structural and magnetic properties thoroughly investigated.
    Each of these materials crystallizes in the $F\bar{4}3m$ space group at ambient temperature. However, in contrast to the end members \ce{GaV4S8} and \ce{GaV4Se8}, that undergo a structural transition to the $R3m$ space group at 42 and 41~K respectively, the solid solutions $(y = 1$ to $7)$ retain cubic symmetry down to 1.5~K. 
    In zero applied  field the end members of the family order ferromagnetically at 13~K (\ce{GaV4S8}) and 18~K (\ce{GaV4Se8}), while the intermediate compounds exhibit a spin-glass-like ground state.
    We demonstrate that the magnetic structure of \ce{GaV4S8} shows localization of spins on the V cations, indicating that a charge ordering mechanism  drives the structural phase transition.   We conclude that the observation of both structural and ferromagnetic transitions in the end members of the series in zero field is a prerequisite for the stabilization of a skyrmion phase, and discuss how the absence of these transitions in the $y = 1$ to $7$ materials can be explained by their structural properties.
\end{abstract}
\begin{document}

\flushbottom
\maketitle
%
%
\thispagestyle{empty}

\section*{Introduction}
The $AB_4X_8$ ($A$ = Ga, Al, Ge, Ti, Fe, Co, Ni, Zn; $B$ = V, Cr, Mo, Re, Nb, Ta; $X$ = S, Se, Te) \cite{barz1973new, perrin1975new, brasen1975magnetic, perrin1976nouveaux, yaich1984nouveaux, rastogi1984electron, johrendt1998crystal, pocha2000electronic, pocha2005crystal, bichler2007tuning, powell2007cation, vaju2008metal, szkoda2009compositional, bichler2011interplay} family of materials adopts a lacunar spinel structure and crystallizes in a non-centrosymmetric, $F\bar{4}3m$, cubic structure at ambient temperature.
These materials exhibit a range of physical properties such as structural phase transitions \cite{powell2007cation, francois1991structural, muller2006magnetic}, ferromagnetism \cite{barz1973new, rastogi1984electron, pocha2000electronic}, antiferromagnetism \cite{johrendt1998crystal, sahoo1993evidence}, superconductivity \cite{pocha2005crystal, abd2004transition}, electric-field-induced resistive switching \cite{vaju2008electric, cario2010Electric}, metal-to-insulator transitions \cite{camjayi2014first}, as well as hosting skyrmions \cite{kezsmarki2015neel, fujima2017thermodynamically, bordacs2017equilibrium}.
In fact, the discovery of skyrmions, nano-sized topologically protected magnetic particle-like spin textures, in \ce{GaV4S8} and \ce{GaV4Se8} has led to intense recent research on lacunar spinels \cite{kezsmarki2015neel, fujima2017thermodynamically, bordacs2017equilibrium, ruff2015multiferroicity, wang2015polar, widmann2017multiferroic, ehlers2016skyrmion, butykai2017characteristics, ehlers2017exchange, ruff2017polar, butykai2017relaxation, reschke2017optical, white2018direct, okamura2019microwave, padmanabhan2019optically}.

Skyrmions attract widespread attention partly as promising candidates for energy-efficient high-density data storage and as spintronic devices \cite{nagaosa2013topological, fert2013skyrmions}. Skyrmions are magnetic vortices that have been observed in magnetic materials in two forms: Bloch skyrmions and N\'{e}el skyrmions, so-named because a cut across the diameter of the skyrmion reveals either a Bloch or N\'{e}el wall of spins. 
Magnetic skyrmions have  been  observed chiefly in chiral B20 and $\beta$-Mn ferromagnets, in the form of a lattice of Bloch-type skyrmions \cite{kanazawa2017noncentrosymmetric}.
In contrast, lattices of  N\' eel-type skyrmions have been found in \ce{GaV4S8} and \ce{GaV4Se8}.
In fact, the N\' eel-type skyrmion lattice in the bulk has, thus far, only been observed in three materials: \ce{GaV4S8}, \ce{GaV4Se8}, and \ce{VOSe2O5} \cite{kurumaji2017neel}.
There is therefore a justifiable interest in enlarging the family of materials hosting the N\' eel-type skyrmion, in order to acquire a better understanding of these spin textures. 
Alteration of the chemical composition, e.g.\ substitution of a transition metal in \ce{Cu2OSeO3}, has proven to be an effective way to expand families of skyrmion-hosting materials and obtain valuable insights into the physics of these systems \cite{stefancic2018origin, wilson2019measuring, birch2019increased}. 
Given the many similarities in the two end members  of the series: \ce{GaV4S8} and \ce{GaV4Se8} (i.e.\  similar structural phase transitions as well as exhibiting ferromagnetic-like ground states and  hosting skyrmion phases), it is interesting to explore the properties of this family further by investigating the intermediate compositions lying between the \ce{GaV4S8} and \ce{GaV4Se8} end members.

Both \ce{GaV4S8} and \ce{GaV4Se8} can be described as cation-deficient $A_{0.5}\square_{0.5}B_2X_4$ spinels, where the symmetry is reduced from $Fd\bar{3}m$ to $F\bar{4}3m$ due to the ordering of the tetrahedral sites.
This results in the shifting of the $B$ atoms from the octahedral voids, forming tetrahedral $B_4$ clusters.
These materials can therefore be represented with the formula $[B_4X_4]AX_4$, where heterocubane-like $[B_4X_4]^{\text{n}+}$ and $[AX_4]^{\text{n}-}$ constituents resemble the arrangement of ions in NaCl \cite{johrendt1998crystal, pocha2000electronic}.
Moreover, both \ce{GaV4S8} and \ce{GaV4Se8} undergo structural phase transitions from high-temperature cubic $(F\bar{4}3m)$ to low-temperature rhombohedral $(R3m)$ symmetry at 42~K (\ce{GaV4S8}) and 41~K (\ce{GaV4Se8}).
The structural phase transition is followed by a magnetic phase transition from paramagnetic to ferromagnetic-like at 13 and 18~K in \ce{GaV4S8} and \ce{GaV4Se8} respectively \cite{pocha2000electronic, bichler2011interplay}.

According to  previous studies, \cite{pocha2000electronic, nakamura2005structural}  \ce{GaV4S8} consists of Ga$^{3+}$ $\left(\text{V}^{3.25+}\right)_4\left(\text{S}^{2-}\right)_8$, where seven electrons in 3d orbitals are involved in V-V metal bonding.
These electrons are, consistent with the cubic ($T_d)$ symmetry, located at the $a_1$ (two electrons), $e$ (four electrons) and $t_2$ (one electron) energy levels.
A Jahn-Teller structural distortion at 42~K drives the material from a cubic to a rhombohedral crystal structure.  This distortion causes a reduction of symmetry to $C_{3V}$, removing the triply-degenerate orbitals at the $t_2$ level, resulting in doubly-degenerate orbitals at the $e$ level (higher in energy) and an orbital at the $a_1$ level (lower in energy) containing one unpaired electron.
In zero magnetic field both \ce{GaV4S8} and \ce{GaV4Se8} undergo a paramagnetic to cycloidal phase transition at 13~K and 18~K respectively\cite{pocha2000electronic, bichler2011interplay}.
Further reduction in temperature for \ce{GaV4S8} yields a transition from cycloidal to a ferromagnetic-like state as opposed to \ce{GaV4Se8} in which a cycloidal state persists down to the lowest measured temperature.
Based on recent muon-spin rotation spectroscopy ($\mu^+$SR) measurements\cite{franke2018magnetic},  differences in the low-temperature magnetic behaviour of \ce{GaV4S8} and \ce{GaV4Se8} are evident. 
In contrast to \ce{GaV4Se8}, \ce{GaV4S8} shows an unusual increase of the local magnetic field with increasing temperature indicating that the magnetic ground states of \ce{GaV4S8} and \ce{GaV4Se8} are significantly different\cite{kezsmarki2015neel,bordacs2017equilibrium}. 
Application of a 30~mT magnetic field to the cycloidal phase in \ce{GaV4S8} and 150~mT in \ce{GaV4Se8} has been found to stabilise N\' eel-type skyrmions.
Moreover, the presence or absence of the skyrmion phase strongly depends on the orientation of applied field with respect to the crystallographic direction, with the most extensive skyrmion phase detected with the field oriented along the [111] crystallographic direction\cite{bordacs2017equilibrium}.

We have undertaken a detailed investigation of the series of compounds obtained by systematically substituting Se at the S site, going from \ce{GaV4S8} to \ce{GaV4Se8}, aimed at understanding  the links between the structural and magnetic properties.
This will help us to explore the possibility of these materials as possible skyrmion hosts.
The synthesized family of polycrystalline GaV$_4$S$_{8-y}$Se$_{y}$ materials were used to grow high-quality single crystals utilizing the chemical vapour transport technique (CVT).
Single crystals and polycrystalline materials were characterized with powder and single crystal X-ray and neutron diffraction, and their magnetic properties were investigated using \textit{ac}/ \textit{dc} magnetometry measurements.
Based on our results we were able to establish magneto-structural relationships that suggest  how the substitution of S with Se affects the magnetic properties and hence the presence or absence of the skyrmion phase in the GaV$_4$S$_{8-y}$Se$_{y}$ materials.   Specifically, we suggest that  charge ordering that likely accompanies a structural distortion in the $y=0$ and $y=1$ materials best explains their ordered magnetic ground states. This structural distortion is not observed in the intermediate $y$ materials, explaining why they form a magnetic spin-glass-like state at low temperature. 

\section*{Results}

\begin{figure}
\includegraphics[width=0.7\linewidth]{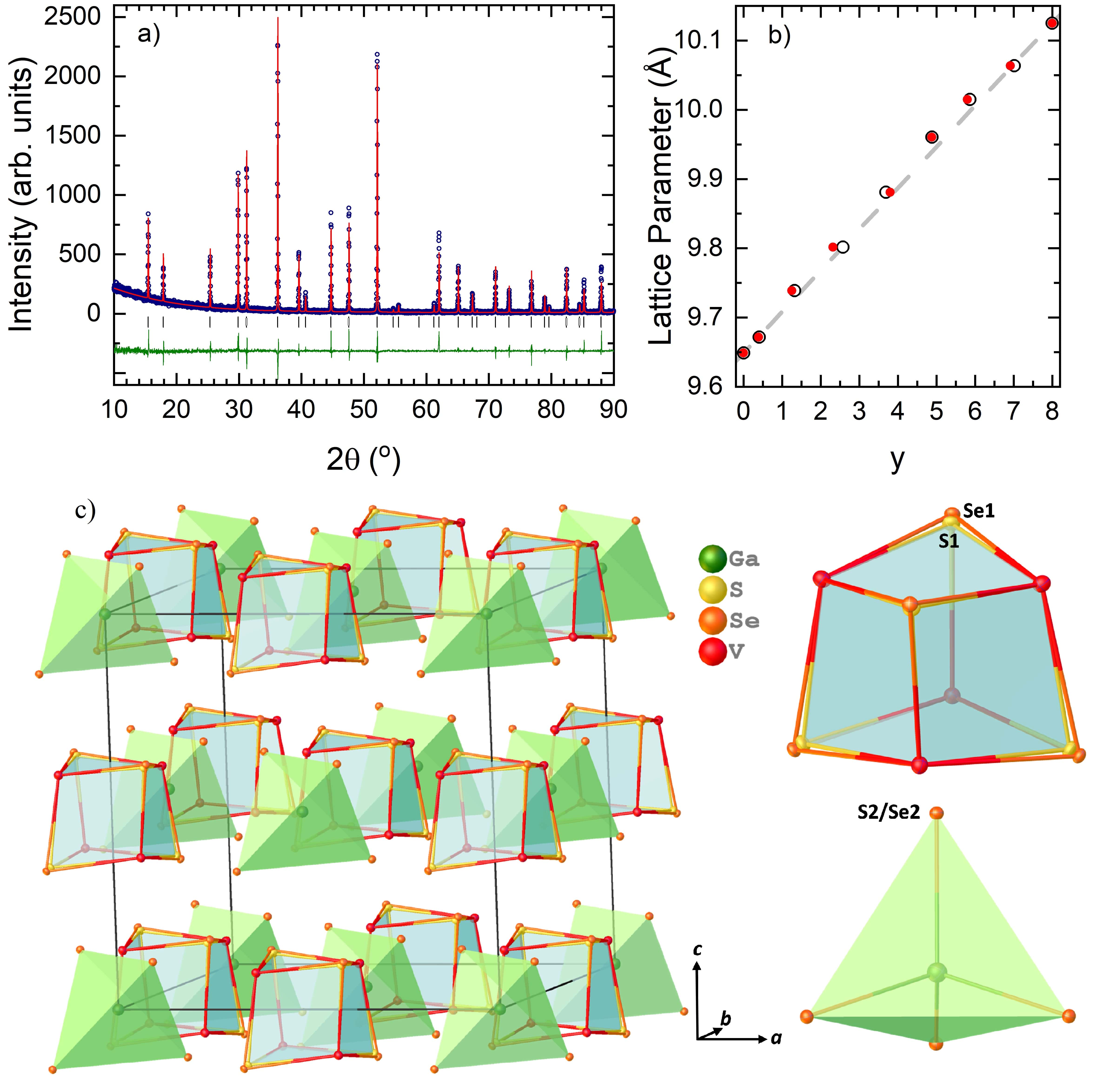}
\caption{a) Powder X-ray diffraction profile of GaV$_4$S$_{4}$Se$_{4}$. 
The experimentally-obtained diffraction profile at ambient temperature (blue open circles), refinement based on the model obtained from single crystal X-ray diffraction at room temperature (red solid line), difference (olive green solid line) and predicted peak positions (black tick marks).
b) Lattice parameters of GaV$_4$S$_{8-y}$Se$_{y}$ obtained from EDX, shown as open black circles, and single crystal X-ray diffraction shown as red circles. The grey dashed line represents Vegard's law between \ce{GaV4S8} and \ce{GaV4Se8}.
The error bars are too small to be seen.
c) Structural model of material \ce{GaV4S4Se4} (refined composition GaV$_4$S$_{4.36}$Se$_{3.64}$) obtained from single crystal X-ray diffraction at 120~K.
The V$_4$S$_{2.85}$/Se$_{1.15}$ heterocubane (top right) and GaS$_{1.51}$Se$_{2.49}$ tetrahedra (bottom right), and their NaCl-like arrangement in the crystal structure (left side).}
\label{Fig:PXRD}
\end{figure}
The phase purity of polycrystalline materials used for the crystal growth was determined by powder X-ray diffraction.
The diffraction profiles of the GaV$_4$S$_{8-y}$Se$_{y}$ samples with $y$ ranging between 0 and 8 could be indexed in a cubic $F\bar{4}3m$ symmetry, as shown in Fig.~\ref{Fig:PXRD}a and Supplementary Figs.~S2 and S3.
The lattice parameter \textit{a} increases monotonically with the increasing \ce{Se} substitution levels, as expected when substituting S$^{2-}$ (ionic radius of 1.70~\AA) with larger Se$^{2-}$ (1.98~\AA) anions (Supplementary Table~S1)\cite{shannon1976revised}.
Materials with $y$ between 0 and 5 are phase pure, while small amounts (up to $3.3\%$) of \ce{VSe2} impurity were observed in the \ce{GaVSSe7} and \ce{GaV4Se8} samples.
In contrast to the rest of the materials investigated, the presence of two cubic phases with slightly different lattice parameters was detected in the GaV$_4$S$_2$Se$_6$ sample.
The refinement of S and Se occupancies revealed higher concentrations of sulphur in GaV$_4$S$_{8-y}$Se$_{y}$ materials than the nominal concentrations used during the synthesis, except in \ce{GaV4S3Se5}.
Moreover, the Se$^{2-}$ is found to preferentially occupy the site near the Ga cations (Ga$X_4$ tetrahedron), while S$^{2-}$ is found to prefer the site near V cations (V$_4X_4$ heterocubane).  

\begin{table}
\centering
\caption{Composition of the single crystals obtained from EDX analysis and from the refinement of single crystal X-ray diffraction (SCXRD) data.}
\label{tab:composition}
\small
    \begin{tabular}{ccc}
        \hline
        Nominal & EDX & SCXD  \\
        \hline
        GaV$_4$S$_8$ & GaV$_4$S$_{7.999(6)}$ & GaV$_4$S$_8$ \\
        GaV$_4$S$_{7}$Se & GaV$_4$S$_{7.591(6)}$Se$_{0.409(6)}$ & GaV$_4$S$_{7.61(2)}$Se$_{0.39(2)}$ \\
        GaV$_4$S$_{6}$Se$_{2}$ & GaV$_4$S$_{6.675(15)}$Se$_{1.325(15)}$ & GaV$_4$S$_{6.748(17)}$Se$_{1.252(17)}$ \\
        GaV$_4$S$_{5}$Se$_{3}$ & GaV$_4$S$_{5.43(5)}$Se$_{2.57(5)}$ & GaV$_4$S$_{5.68(2)}$Se$_{2.32(2)}$ \\
        GaV$_4$S$_{4}$Se$_{4}$ & GaV$_4$S$_{4.306(8)}$Se$_{3.694(8)}$ & GaV$_4$S$_{4.20(3)}$Se$_{3.80(3)}$ \\
        GaV$_4$S$_{3}$Se$_{5}$ & GaV$_4$S$_{3.118(14)}$Se$_{4.882(14)}$ & GaV$_4$S$_{3.12(4)}$Se$_{4.88(4)}$ \\
        GaV$_4$S$_{2}$Se$_{6}$ & GaV$_4$S$_{2.135(15)}$Se$_{5.865(15)}$ & GaV$_4$S$_{2.20(5)}$Se$_{5.80(5)}$ \\
        GaV$_4$SSe$_{7}$ & GaV$_4$S$_{0.98(2)}$Se$_{7.02(2)}$ & GaV$_4$S$_{1.09(5)}$Se$_{6.91(5)}$ \\
        GaV$_4$Se$_8$ & GaV$_4$Se$_{8.00(3)}$ & GaV$_4$Se$_8$\\
        \hline
    \end{tabular}
\end{table}

The compositions of single crystals with $y$ between 0 and 8 were determined by refining single crystal X-ray diffraction data, and were further checked using energy dispersive x-ray measurements as shown in Table~\ref{tab:composition}.
The obtained lattice parameters can be seen to follow Vegard's law (Fig.~\ref{Fig:PXRD}b).
Single crystal X-ray diffraction shows that GaV$_4$S$_{8-y}$Se$_{y}$ $(y = 0$ to $8)$ adopts the same crystal structure down to 50~K, where Ga$X_4$ and V$_4X_4$ units exhibit the NaCl-like arrangement as shown in Fig.~\ref{Fig:PXRD}c.
In contrast to the two end compounds, \ce{GaV4S8} and \ce{GaV4Se8}, substitutional disorder is observed in all the intermediate solid solutions examined.
To model the substitutional disorder properly, different refinement approaches were used: either allowing the atomic positions of S and Se to refine separately or constraining them.
The most reliable structural models were obtained when S2 and Se2 in the Ga$X_4$ tetrahedra were fixed to occupy the same atomic position, and the atomic positions of S1 and Se1 in the V$_4X_4$ unit were allowed to refine separately, but were constrained by the crystal symmetry.
The occupancies of sulphur and selenium were allowed to be freely refined, with the total occupancy constrained to 1, and there is clear evidence that S prefers to occupy the atomic position in the V$_4X_4$ cluster, while the Se prefers to be situated in the Ga$X_4$ tetrahedron (Supplementary Table~S2 and S3).
These findings are slightly at odds with the hard-soft acids-bases \cite{pearson1963hard} concept, where Ga$^{3+}$ is classified as a harder acid than V$^{3+}$, and S$^{2-}$ is classified as a stronger base than Se$^{2-}$, and therefore it would be expected that Se$^{2-}$ would prefer to bind to V$^{3+}$, and S$^{2-}$ to bind to Ga$^{3+}$.
However, Se$^{2-}$ is considerably bigger than S$^{2-}$, hence the tetrahedral coordination in the Ga$X_4$ tetrahedron is preferred over the trigonal coordination in the V$_4X_4$ cluster.

\begin{figure}
\includegraphics[width=\linewidth]{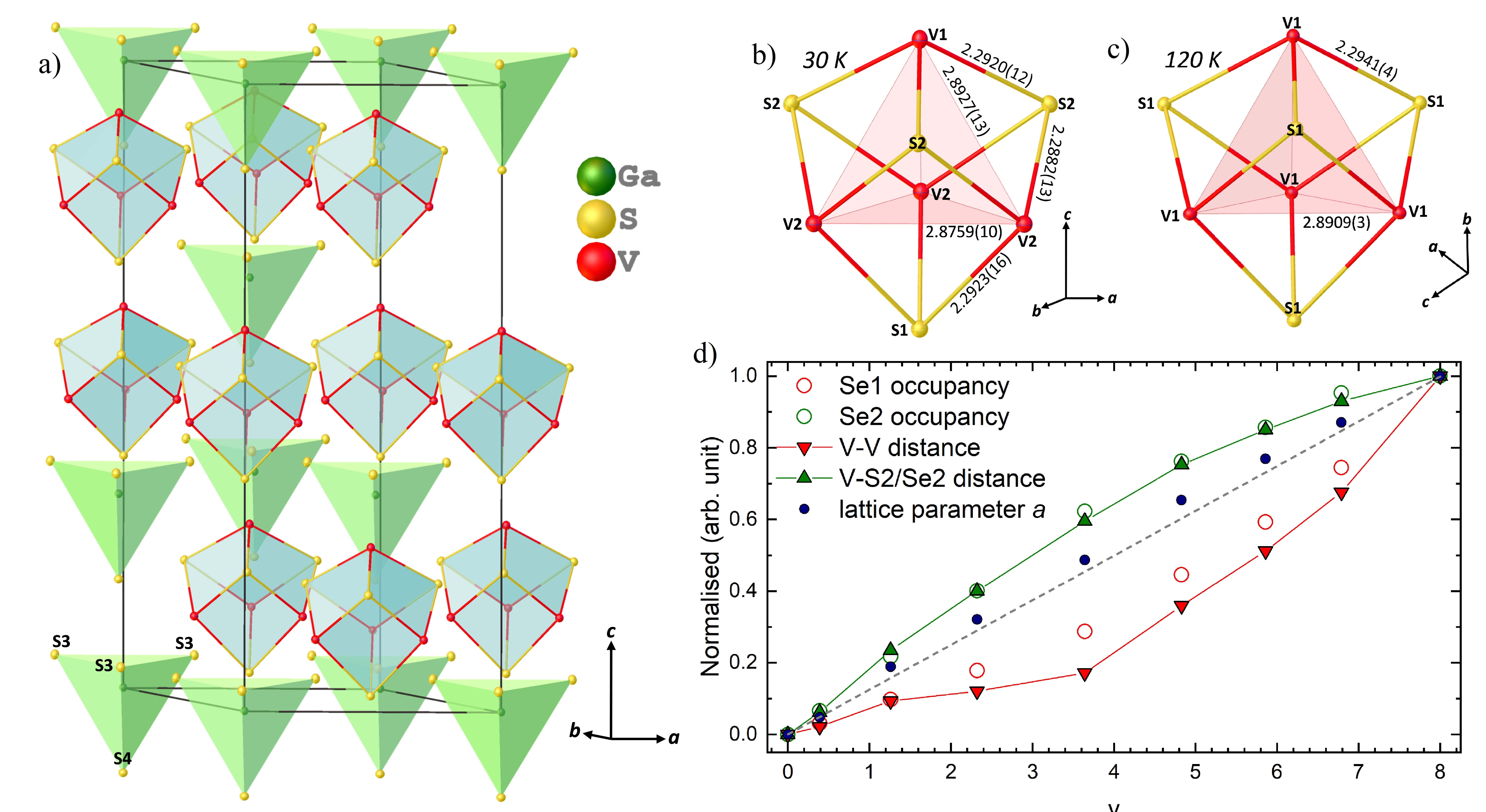}
\caption{a) Structural  model  of  material \ce{GaV4S8} obtained  from the data collected on  single  crystal  X-ray  diffraction  at  30~K.
b) The distorted and c) not distorted \ce{V4S4} heterocubane like entity based on the data collected at 30 and 120~K respectively. 
d) Correlation between V-V (red triangles), V-S1/Se1 (green triangles) bond distances and Se2 (red circles), Se1 (green circles) site occupancies and the lattice parameters $a$ (blue circles) obtained from single crystal X-ray diffraction on a laboratory diffractometer.
The grey dotted line represents the idealized Vegard's law.
}
\label{Fig:bond_dist}
\end{figure}
As mentioned above, a linear increase of the lattice parameter \textit{a} was observed and the obtained values are seen to follow Vegard's law \cite{vegard1921konstitution}, depicted as the grey dotted line in Fig.~\ref{Fig:bond_dist}d. 
The inter-constituent (Ga$X_4$ and V$_4X_4$) distances and the Ga-S2/Se2 distances increase linearly, following the same trend as the lattice parameters.
However, the intra V$_4X_4$ bond distances, angles and tetrahedron (V$_4$, S1$_4$ and Se1$_4$) volumes in the intermediate compounds ($y$ between 1 and 7) do not show this trend (Supplementary Table~S3).
A close examination of the structural models reveals that V-V distances within the cluster increase much more slowly than expected as a function of $y$, while V-S2/Se2 distances increase slightly faster.
Moreover, a higher deviation from the linearity in V-V distance can be observed in the material $y = 4$.
This can be partially explained by the S1 and Se1 site occupancies. However, the trend of the Se1 site occupancy does not exactly follow the trend of the V-V bond distances, as shown in Fig.~\ref{Fig:bond_dist}d.
In contrast, the changes of the V-S2/Se2 bond distances are in very good agreement with changes of Se2 site occupancy.
These findings indicate that the tendency of S$^{2-}$ and Se$^{2-}$ to occupy the atomic positions in the V$_4X_4$ cluster and in the Ga$X_4$ tetrahedron respectively is the driving force for the structural motifs observed in the GaV$_4$S$_{8-y}$Se$_y$ solid solutions.
Additionally, the volumes of the Ga$X_4$ and V$_4$ tetrahedra increase with increasing Se substitution levels (as expected, the volume of V$_4$ tetrahedra follow the same trend as V-V bond distances), however, the volume of (S1)$_4$ and (Se1)$_4$ tetrahedra stays nearly constant throughout the series (Fig.~S4).            

The end compounds, \ce{GaV4S8} and \ce{GaV4Se8}, undergo a structural phase transition at 42~K and 41~K respectively.
At this transition, one of the vanadium and one of the sulphur/selenium atoms move along the [111] crystallographic direction, as shown in Fig.~\ref{Fig:bond_dist}a, reducing the $F\bar{4}3m$ to $R3m$ symmetry, where the cubic diagonal becomes the $c$ crystallographic lattice.
Lattice parameters of $a = b = 6.8098(3)$~\AA \ and $c = 16.7100(17)$~\AA \ were obtained for \ce{GaV4S8} at 30~K, below the structural transition, and are in good agreement with the reported values \cite{pocha2000electronic}.
Three of the V1-V2 bond distances $(2.8927(13)~\text{\AA})$ are elongated in comparison to the three V2-V2 distances $(2.8759(10)~\text{\AA})$ and the rhombohedral angle of $59.62(3)^\circ$ deviates from $60^\circ$ observed in cubic symmetry; see Fig.~\ref{Fig:bond_dist}.
Also, three V1-S2 $(2.2920(12)~\text{\AA})$ and three V2-S1 $(2.2923(16)~\text{\AA})$ bond distances are longer than the remaining six V2-S2 $(2.2882(13)~\text{\AA})$ bond distances as shown in Fig.~\ref{Fig:bond_dist}.
Additionally, the \ce{GaS4} tetrahedron distorts along the crystallographic $c$  direction with one Ga-S4 $(2.271(2)~\text{\AA})$ bond distance elongated compared to the remaining three Ga-S3 $(2.2664(10)~\text{\AA})$ bond distances (Fig.~\ref{Fig:bond_dist}).
The observed structural distortion is in very good agreement with that reported previously\cite{pocha2000electronic, powell2007cation}.
However, the changes in bond distances obtained in single crystals are less pronounced than those seen in polycrystalline samples.

\subsection*{Neutron Diffraction}
\begin{figure}
\includegraphics[width=0.7\linewidth]{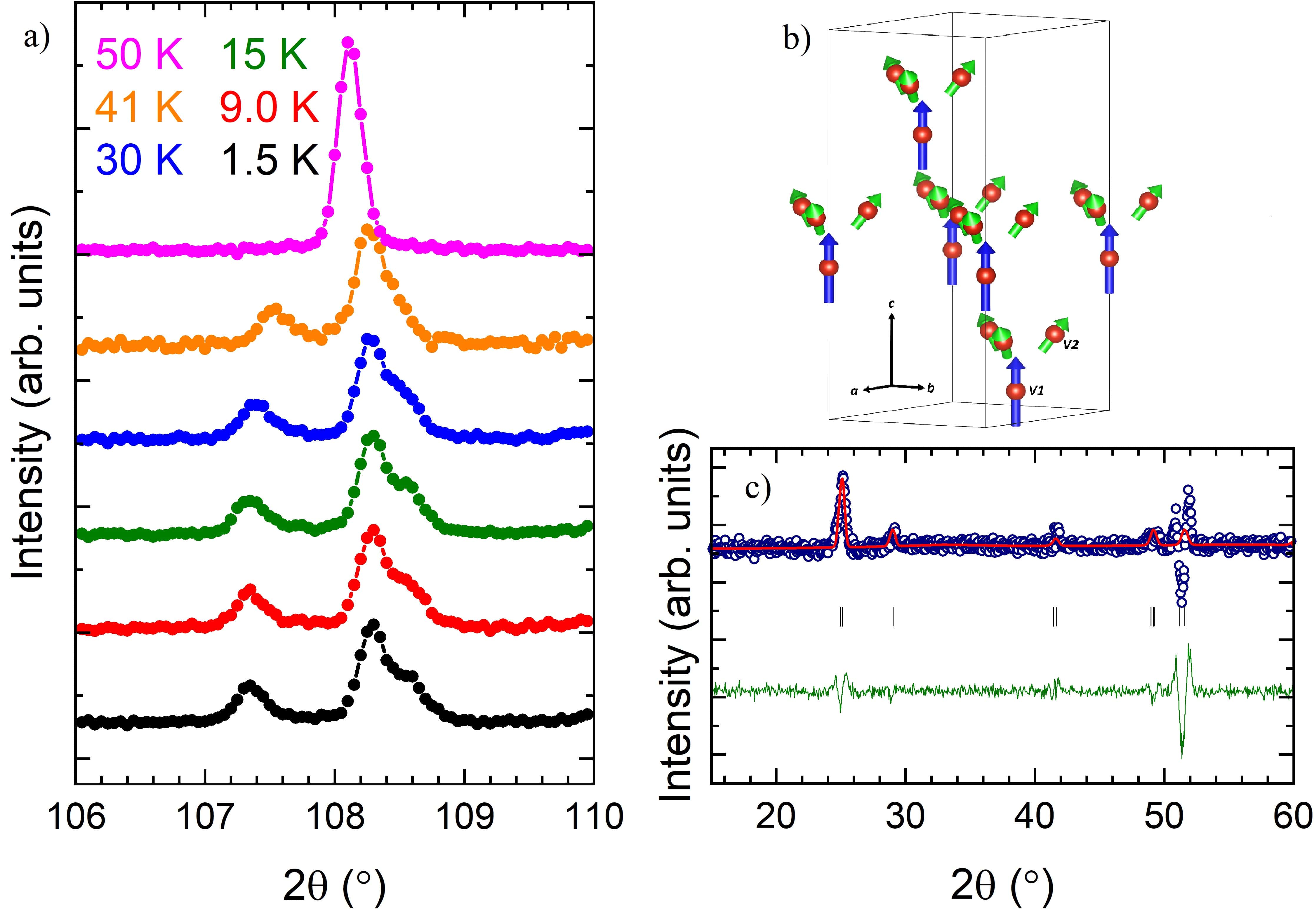}
\caption{a) Powder neutron diffraction profiles of \ce{GaV4S8} measured on the D2B diffractometer at the ILL, showing the splitting of the cubic $(448)$ peak below the structural phase transition of 42~K indicating a structural phase transition from a $F\bar{4}3m$ to $R3m$ structure.
The black solids circles correspond to data taken at 1.5~K, red at 9~K, olive at 15~K, blue at 30~K, orange at 41~K and magenta at 50~K.
b) Magnetic difference pattern of GaV$_4$S$_{8}$ at $1.5-15$~K. 
The experimentally-obtained diffraction profile on a powder sample, using the  D20 diffractometer at the ILL (blue open circles), refinement based on the commensurate $R3m'$ model (red solid line), difference (olive green solid line) and predicted peak positions (black tick marks). c) Magnetic model obtained from refinement of the magnetic difference pattern in the $R3m'$ Shubnikov group, where the magnetic moments on V1 atoms are shown as blue arrows and those on V2 atoms are shown as green arrows.}
\label{Fig:D2B_peaksplit}
\end{figure}

To complement the X-ray diffraction data, powder neutron diffraction was performed on the end compound (\ce{GaV4S8}) and one intermediate solid solution (\ce{GaV4S6Se2}) with representative properties of the solid solutions, to obtain the nuclear structures at a range of temperatures.
Powder neutron diffraction data obtained on \ce{GaV4S8} 
measured at 50~K, above the structural transition, refine in the high temperature $F\Bar{4}3m$ space group (Supplementary Table~S4).
Due to the small coherent scattering cross section of V, the positions of the V atoms were constrained to those obtained in single crystal X-ray diffraction.
The evidence of a structural phase transition from cubic to a pseudo-cubic rhombohedral structure can be clearly seen through the splitting of some of the cubic Bragg peaks such as the cubic $(448)$ peak as shown in Fig.~\ref{Fig:D2B_peaksplit}a.
Below the structural phase transition at 41~K, the crystal structure remains the same until 1.5~K and refines in the $R3m$ space group, as shown in Supplementary Table~S4.
These results are in good agreement with those reported in the literature \cite{pocha2000electronic}.
Powder neutron diffraction data 
for \ce{GaV4S6Se2} show the absence of a structural phase transition, with data refining in a $F\bar{4}3m$ space group down to 1.5~K, as shown in Supplementary Fig.~S5.
The site location and occupancy of S$^{2-}$ and Se$^{2-}$ within the V$_4X_4$ units are in good agreement with X-ray diffraction results.
Similar to the X-ray diffraction data, S1 and Se1 are found to occupy different sites within the \ce{V4X4} cluster with a higher fraction of S$^{2-}$ located within the V$_4X_4$ cluster and Se$^{2-}$ preferentially occupying sites in the Ga$X_4$ tetrahedron.

Similar to powder neutron diffraction, one single crystal of the end composition (\ce{GaV4Se8}) and one single crystal of a member of the intermediate materials (\ce{GaV4S4Se4}) with indicative properties of the set of intermediate solid solutions were chosen for investigation.
The data obtained for GaV$_4$Se$_{8}$ refines in a $F\Bar{4}3m$ space group at 55~K whereas the data between 30 to 1.5~K refine in a $R3m$ space group, in very good agreement with both our X-ray data and the literature\cite{bichler2011interplay} (see Supplementary Table.~S6).
As in the analysis of the powder neutron diffraction, the positions of the vanadium atoms were constrained to those obtained from single crystal X-ray diffraction.
The low temperature measurements show that \ce{GaV4S4Se4} does not undergo a structural phase transition, and remains cubic down to 1.5~K, similar to all the other intermediate compositions examined.
The nominal \ce{GaV4S4Se4} single crystal is found to be GaV$_4$S$_{4.36}$Se$_{3.64}$,based on the refinement of the S and Se site occupancies, which is in excellent agreement with single crystal x-ray diffraction (SCXRD) and EDX (Table.~\ref{tab:composition}).
Moreover, the individual site occupancies show that there is a higher fraction of  S atoms situated within the V$_4X_4$ cluster and a lower fraction in the Ga$_4X_4$ tetrahedron, also in good agreement with our powder neutron diffraction and X-ray diffraction data.

\subsection*{Magnetic Neutron Scattering}
Magnetic structures for the end compound \ce{GaV4S8} and two representative members of the intermediate solid solutions (\ce{GaV4S6Se2} and \ce{GaV4S4Se4}) were probed using both powder and single crystal diffraction.
The magnetic scattering was isolated by taking the difference pattern between 1.5 and 15~K, above and below the magnetic transition. 
For \ce{GaV4S8}, the magnetic peaks are located at the same positions as the nuclear Bragg peak positions and have a symmetrical peak shape indicating a $k=0$ propagation vector as seen in Fig.~\ref{Fig:D2B_peaksplit}c.
Magnetic symmetry analysis with help of the program BASIREPS \cite{rodriguez2010program, ritter2011neutrons} was used to determine the allowed irreducible representations (IR) and their basis vectors (BV) for the two vanadium sites.
Testing the different allowed IRs showed that the magnetic difference pattern for \ce{GaV4S8} refines best in a $R3m$' magnetic structure, shown in Fig.~\ref{Fig:D2B_peaksplit}b.
The V1 atoms refine to have a magnetic moment of 0.30(6)~$\mu_\text{B}$, pointing along the hexagonal $c$ axis whereas the V2 magnetic moments have a smaller moment of 0.18(7)~$\mu_\text{B}$ and are canted outward from the cluster.
The localization of a higher moment on the distorted V1 site is consistent with density functional theory (DFT) calculations showing a higher spin density residing on the distorted V1 atom as opposed to the V2 atoms \cite{muller2006magnetic, nikolaev2019microscopic}.
It is worth noting that, due to the nature of powder neutron diffraction, it cannot be established if the orientations of the moments are in the configuration shown in Fig.~\ref{Fig:D2B_peaksplit}b or pointing in the opposite direction with the same magnitude, i.e.\ moments on the V2 atoms pointing in towards the cluster with the moment on the V1 atom pointing in the negative crystallographic $c$ axis.
Attempts were made to refine the structure with only one magnetic moment per \ce{V4} cluster to be localized either on the V1 site or in the centre of the cluster. However, a single magnetic moment per \ce{V4} cluster does not produce the correct ratios of peak intensities, supporting the fact that both V1 and V2 carry a magnetic moment.

Powder neutron diffraction patterns of \ce{GaV4S6Se2} show no difference between the diffractograms measured at 1.5 and 15~K, as shown in Supplementary Fig.~S5, confirming the absence of any magnetic intensity.
Additionally, single crystal neutron diffraction of \ce{GaV4S4Se4} 
revealed no magnetic scattering intensity at 1.5~K.
There is a strong indication that no long-range magnetic ordering is present in solid solutions based on our results obtained on the two representative materials.

\begin{figure}
\centering
\includegraphics[width=0.7\linewidth]{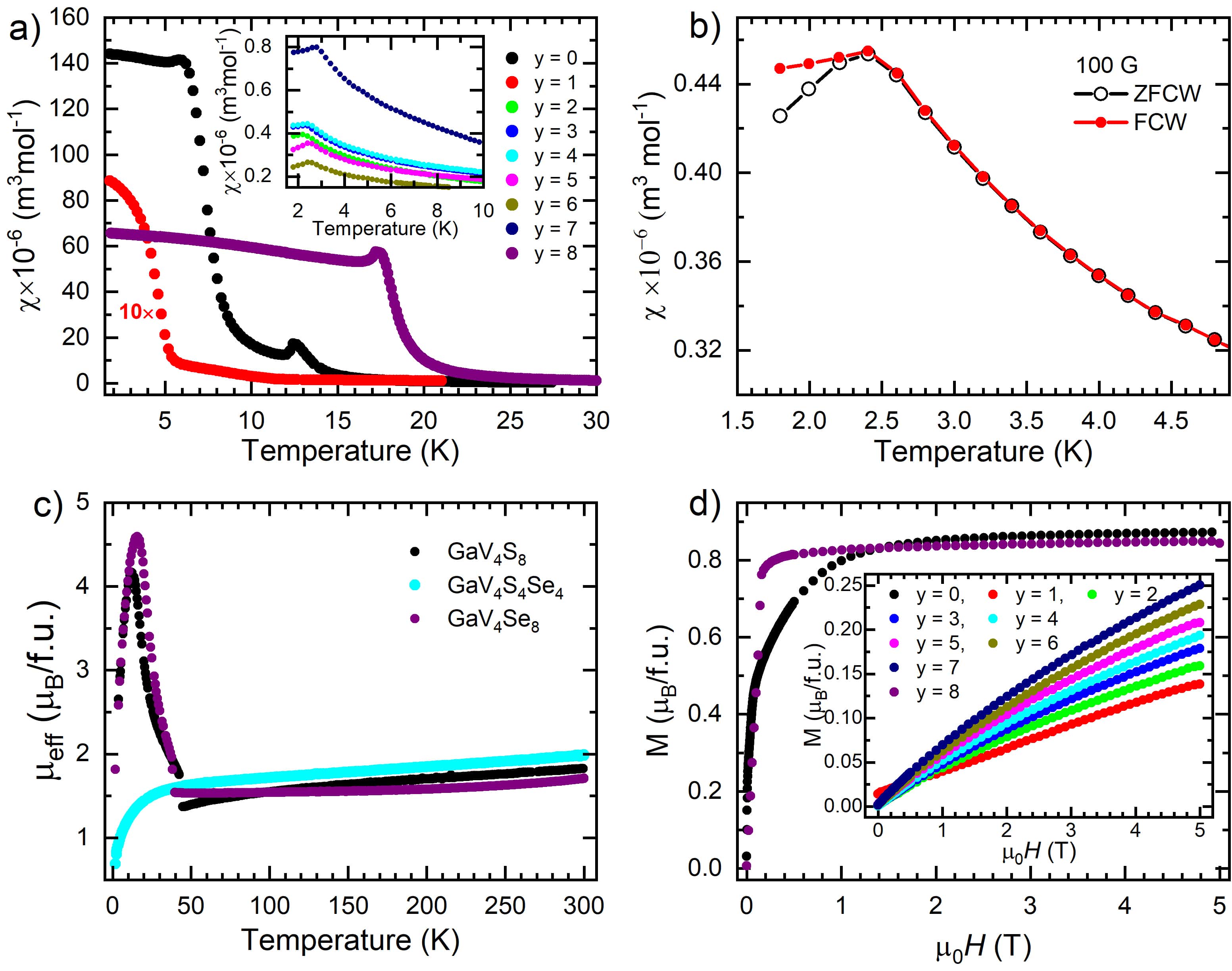}
\caption{a) Magnetic susceptibility vs temperature for the GaV$_4$S$_{8-y}$Se$_y$ family measured in the applied magnetic field of 10~mT.
b) Bifurcation of zero-field cooled warming, shown as open black circles, and field cooled warming, shown as closed red circles for \ce{GaV4S4Se4}.
c) Evolution of the effective moment $\mu_{\text{eff}}$ with temperature for \ce{GaV4S8}, \ce{GaV4S4Se4}, and \ce{GaV4Se8} single crystals calculated from magnetization measurements.
d) Magnetic isotherms to investigate the saturation magnetization at 1.8~K of the GaV$_4$S$_{8-y}$Se$_y$ in an applied magnetic field $\mu_0H$.
}
\label{Fig:mag_mes}
\end{figure}

\begin{figure}
\centering
\includegraphics[width=0.7\linewidth]{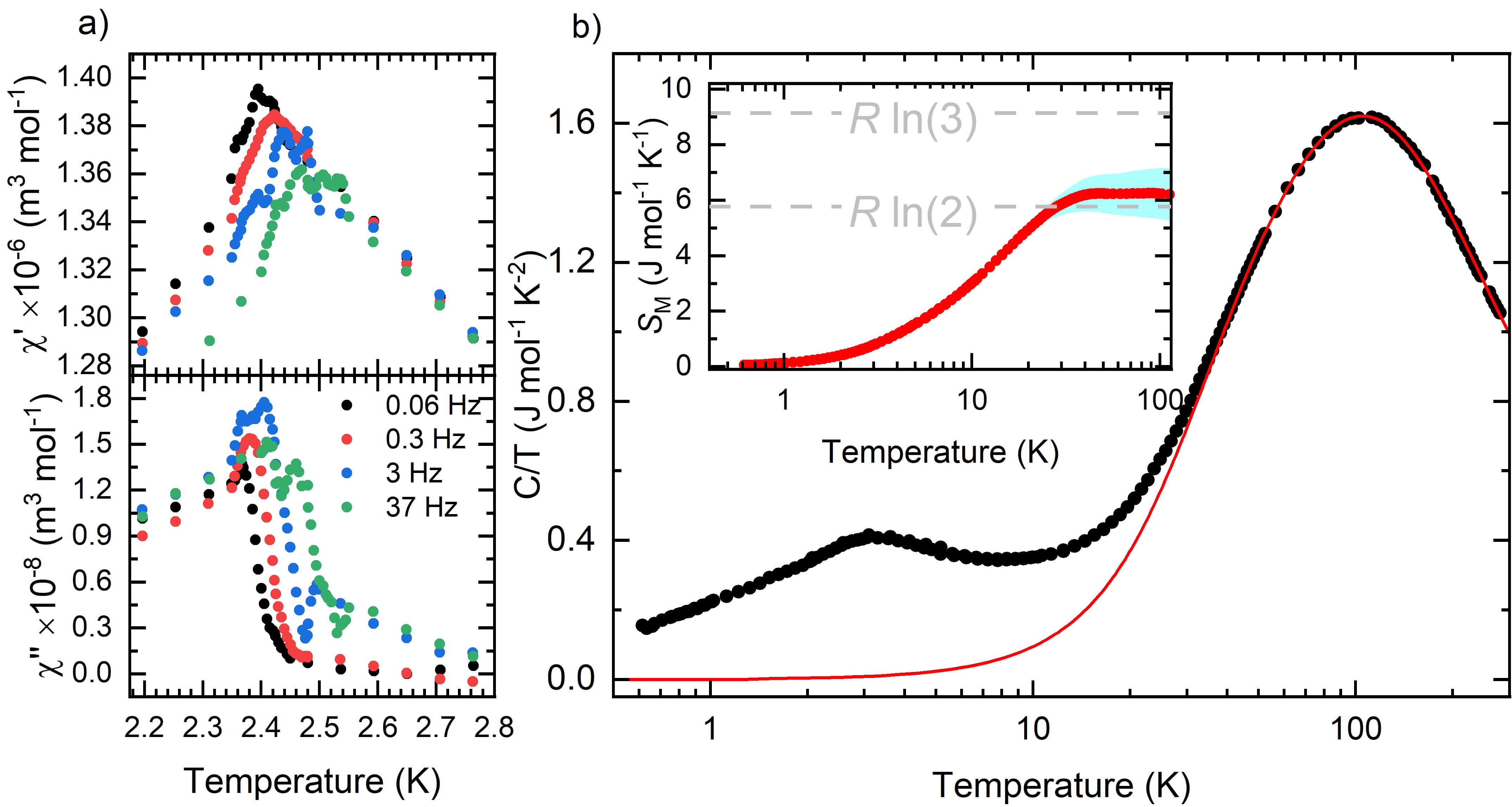}
\caption{a) Real $\chi'$ and imaginary $\chi''$ component of the \textit{ac} susceptibility on \ce{GaV4S4Se4} single crystals spanning the freezing transition $T_f$ for a range of frequencies.
b) Temperature dependence of the specific heat in zero field for \ce{GaV4S4Se4} plotted on a semi-logarithmic scale for clarity.
The phonon contribution is shown as a solid red line.
The inset shows the magnetic entropy recovered after subtracting the phononic contribution from the specific heat. The $1\sigma$ error band is shown in blue.
The grey dashed lines in the inset show the values for $R\ln{2}$ and $R\ln{3}$ corresponding to the theoretical entropy of a spin 1/2 and spin 1 system respectively.
}
\label{Fig:mag}
\end{figure}

\subsection*{Magnetic Measurements}
Bulk magnetization measurements on single crystals were carried out to identify the nature of magnetic ordering occurring within the GaV$_4$S$_{8-y}$Se$_y$ family of materials.
Obvious differences in the magnetic susceptibilities across the GaV$_4$S$_{8-y}$Se$_{y}$ family are observed, as shown in Fig.~\ref{Fig:mag_mes}a.
The end compounds, \ce{GaV4S8} and \ce{GaV4Se8}, undergo long-range ferromagnetic-like ordering at 13 and 18~K, respectively, with \ce{GaV4S8} having a susceptibility over twice as large as its selenium counterpart \ce{GaV4Se8} \cite{pocha2000electronic, bichler2011interplay}.
The structural phase transitions can be observed as large discontinuities in the inverse magnetic susceptibilities, as shown in Supplementary Fig.~S6, as was  previously reported in the powder materials\cite{pocha2000electronic, bichler2007tuning, widmann2017multiferroic, nakamura2005structural, powell2007cation, bichler2011interplay}.
This highlights the magneto-structural link between the cubic and rhombehedral phase, with the discontinuities likely related to the lifting of the degeneracy of V orbitals from the structural phase transition leading to a change in the location of spin density and hence magnetic interactions.
The effective moment calculated from the susceptibility by using the Curie law $\mu_\text{eff}=797.8\sqrt{\chi_m T}$ is given in  Fig.~\ref{Fig:mag_mes}c.
A discontinuity in the effective moment $\mu_\text{eff}$ can be seen at the temperature of the structural phase transition in \ce{GaV4S8} (42~K) and \ce{GaV4Se8} (41~K), and a large increase in $\mu_\text{eff}$ is seen with decreasing temperature.
The peak in effective moment is in good agreement with the onset of long-range magnetic order.
Above the structural phase transition in \ce{GaV4S8} and \ce{GaV4Se8}, $\mu_\text{eff}$ continuously increases in the temperature region between $\sim 50$ and $300$~K, as previously reported \cite{pocha2000electronic}.
This means that the Weiss temperature cannot be accurately extracted from $\chi$ vs $T$ measurements, although it can be seen that the interactions change from antiferromagnetic to ferromagnetic at the structural phase transition.
Based on the $M$ vs\ $\mu_0 H$ measurements shown in Fig.~\ref{Fig:mag_mes}d, the saturation magnetization at applied fields of 5~T and 1.8~K was found to be 0.87(2)~$\mu_\text{B}/$f.u. and 0.85(2)~$\mu_\text{B}/$f.u. for \ce{GaV4S8} and \ce{GaV4Se8}, respectively.
These values are in good agreement with those reported previously \cite{bichler2007tuning}.
Additionally, the value of 0.87(2)~$\mu_\text{B}/$f.u. for \ce{GaV4S8} is also in good agreement with the value of $0.84(7)~\mu_\text{B}/$f.u., the total magnetic moment per \ce{V4} cluster, obtained from our neutron diffraction refinements.

A large suppression in both the magnetic transition temperature and the magnitude of the magnetization is evident in all of the intermediate compounds measured, as seen in Fig.~\ref{Fig:mag_mes}a, suggesting that the absence of a structural phase transition has a dramatic effect on the magnetism.
Examination of the magnetization at low temperature shows a bifurcation of the zero-field-cooled-warming (ZFCW) and field-cooled-warming (FCW) susceptibility curves, shown in Fig.~\ref{Fig:mag_mes}b, indicative of either a canted antiferromagnetic ordering or a glassy transition. It is notable that similar behaviour has been seen in related compounds\cite{bichler2007tuning,powell2007cation,vaju2008metal,szkoda2009compositional,bichler2011interplay}.
Our recent $\mu^+$SR measurements show no evidence for long-range magnetic ordering in the intermediate compounds and instead indicate the presence of a glass-like state \cite{franke2018magnetic}.  
In Fig.~\ref{Fig:mag_mes}c, it can be seen that even though \ce{GaV4S4Se4} does not undergo a structural phase transition, the effective moment starts to decrease rapidly at a similar temperature to where the structural phase transitions occur in \ce{GaV4S8} and \ce{GaV4Se8}, suggesting the onset of magnetic correlations. The effective moment behaves qualitatively very differently with temperature in the intermediate compounds, consistent 
with a lack of ordered magnetic response and 
the likely formation of a low-temperature glassy state.

In order to investigate the magnetic behaviour of one of the intermediate compositions in depth, \textit{ac} susceptibility measurements on \ce{GaV4S4Se4} single crystals were performed.
Fig.~\ref{Fig:mag}a shows behaviour again indicative of a spin glass, with the peak of the in-phase and out-of-phase components of susceptibility shifting in temperature and amplitude, depending on the applied \textit{ac} frequency.
The peaks observed in the \textit{ac} susceptibility data are caused by a freezing of the spins in the material, and hence correspond to a freezing temperature $T_f$.
The freezing temperature for \ce{GaV4S4Se4} can be seen to be $\sim 2.4$~K.
The frequency dependence of the peaks in $\chi'_{ac}$ can be characterized by a shift of the freezing temperature per decade of frequency given by
\begin{equation}
    \delta T_f = \frac{\Delta T_f}{T_f \Delta(\log_{10}\nu)},
\end{equation}
where $\nu$ is the frequency of the applied \textit{ac} magnetic field.
For \ce{GaV4S4Se4}, $\delta T_f$ is found to be $\sim 0.015$.
This shift is in good agreement with those observed for canonical spin-glasses \cite{mydosh}.

\subsection*{Heat Capacity}
Heat capacity measurements on polycrystalline \ce{GaV4S4Se4} were performed to confirm the absence of a structural phase transition and investigate the onset of magnetic order in the intermediate compositions.
Unlike the \ce{GaV4S8} and \ce{GaV4Se8} \cite{widmann2017multiferroic, ruff2017polar}, no sharp lambda anomaly, indicative of the structural phase transition, can been seen for \ce{GaV4S4Se4}, as shown in Fig.~\ref{Fig:mag}b.
Moreover, at low temperatures a broad peak can be seen centered at $\sim3.1$~K.
The position of this peak is located at $\sim 20\%$ above $T_f$, inline with known spin glass materials \cite{mydosh}, and indicates the slowing down and freezing of magnetic moments above $T_f$.

The phonon contributions were estimated using one Debye and three Einstein modes.
Fitting was performed in the temperature region between 45 to 300~K with a Debye temperature of 201(4)~K, and three Einstein frequencies corresponding to temperatures of 263(8), 465(8), and 508(13)~K.
By subtracting the phonon contribution, the associated magnetic entropy can be calculated as shown in the inset of Fig.~\ref{Fig:mag}b.
The magnetic entropy can be seen to be recovered up to approximately 42~K where the entropy is in good agreement with the value of $R\ln{2}$, the entropy corresponding to a spin 1/2 system.
These heat capacity results are consistent with those of magnetometry, indicating that there are short-range magnetic correlations occurring at temperatures far above the spin glass transition.
The phonon subtracted heat capacity below $T_f$ shows $T^\frac{3}{2}$ like dependence deviating from the linear expected behaviour of a canonical spin glass.

\section*{Discussion}
Our structural analysis of single crystals confirms that the end compounds, \ce{GaV4S8} and \ce{GaV4Se8}, undergo a structural phase transition from cubic ($F\bar{4}3m$) to rhombohedral ($R3m$) symmetry at 42 and 41~K respectively, as previously reported \cite{pocha2000electronic, bichler2011interplay}.
The vanadium and $X$ atoms in the V$_4X_4$ ($X$ = S or Se) heterocubane-like entity face along all four [111] crystallographic directions in the cubic structure. 
At the structural phase transition one of the vanadium and the $X$ pairs of atoms distorts, elongating along the cubic diagonal,  reducing the symmetry from cubic to rhombohedral, as shown in Fig. \ref{Fig:bond_dist}.
This distortion has an equal probability to occur along any of the four [111] crystallographic directions.
(In addition, our powder neutron diffraction measurements indicate the presence of noticeable negative thermal expansion in \ce{GaV4S8} in the rhombohedral phase (Table~S4). 
In the temperature region between 30 and 9~K, the lattice parameter \textit{c} increases by 0.04\%, while the lattice parameter \textit{a} contracts by 0.01\%, resulting in a negative thermal expansion.
The origin of this subtle negative thermal expansion is a subject of further investigation.)

The structural phase transition is followed by long-range ferromagnetic-like ordering at 13~K and 18~K for \ce{GaV4S8} and \ce{GaV4Se8} respectively.
Our powder neutron diffraction measurements allowed the determination of the magnetic structure of \ce{GaV4S8} revealing that, on the neutron time scale,  spins are localized on the individual vanadium cations.
A magnetic moment of $0.30(6)~\mu_\text{B}$ on V1 points along the $c$ axis, while a V2 cation moment  of magnitude $0.18(7)~\mu_\text{B}$  is canted outward from the \ce{V4} cluster (Fig.~\ref{Fig:D2B_peaksplit}b), yielding a magnetic moment of $0.84(7)~\mu_\text{B}$ in the spin-polarized state.
This value is in very good agreement with the magnetic moment of $0.87~\mu_\text{B}$  obtained from our magnetization measurements ($M$ vs $\mu_0 H$ at 1.8~K) and with previously-reported values \cite{pocha2000electronic, bichler2007tuning}. The spin arrangement on V cations in the \ce{V4S4} cluster observed in our magnetic model is similar to the arrangement of spin density on V cations in \ce{GaV4S8} obtained from DFT calculations \cite{nikolaev2019microscopic, zhang2017magnetic, muller2006magnetic}.

To date, the electronic state and the origin of long-range ferromagnetic-like ordering in \ce{GaV4S8} and \ce{GaV4Se8} have been explained via a recombination of $d$ orbitals on the V cations, leading to the formation of a molecular system with $S = 1/2$ per \ce{V4} cluster. In this picture the structural phase transition is driven by the Jahn-Teller effect, distorting one of the V and S atoms from an ideal cubic symmetry \cite{pocha2000electronic, nakamura2005structural}.
This model has been used to successfully explain a range of physical properties in this series, such as the polar relaxation dynamics observed in dielectric and THz spectroscopy \cite{wang2015polar}, whose characteristic behaviour can be described within the framework.
However, it is hard to reconcile this model with the magnetic behaviour of these materials. We note that the experimentally-obtained magnetic moment $(0.821~\mu_\text{B})$ is considerably lower than the $1~\mu_\text{B}$ anticipated for a $S=1/2$ system.
Additionally, the ionic radius of the V$^{3+}$ cation is 0.64~\AA \ \cite{shannon1976revised}, and so the bond distances between V cations within the cluster ($\sim 2.9$~\AA) are exceptionally long to permit a significant overlap of $d$ orbitals, required for this model.
Similarly, direct exchange interactions over the inter-cluster distance of $\sim 3.9$~\AA \ are unlikely.

Here we present an alternative interpretation.
It is notable that both intra- and inter-cluster V cations are bonded to bridging S anions with an angle of $\sim 102^\circ$.
This configuration permits ferromagnetic superexchange interactions between spins located on the V cations via S$^{2-}$ or Se$^{2-}$ \cite{goodenough1955theory, kanamori1959superexchange}.
We propose, therefore, that the low-temperature electronic state (and consequently the origin of long-range ferromagnetic ordering in \ce{GaV4S8}) 
follows from the structural phase transition at 42~K, where there is a freezing out of a delocalized electron  onto one of the vanadium atoms, forming three V$^{3+}$ and one V$^{4+}$ cations (i.e.\ a charge-ordered state). Evidence for the presence of V$^{3+}$ and V$^{4+}$ in the ratio of 0.75:0.25 in both cubic and rhombohedral structures comes from bond-valence analysis performed using the SoftB software \cite{chen2019softbv}.
Charge ordering is also consistent with a sharp increase of the resistivity at the structural phase transition observed previously\cite{pocha2000electronic, widmann2017multiferroic}.

Compared to the end compounds (\ce{GaV4S8} and \ce{GaV4Se8}), the solid solutions (with $y$ between 1 and 7) show quite distinct behaviour.
Substitutional disorder is found to be present in all intermediate compounds (shown in Fig.~\ref{Fig:PXRD}~c), where Se prefers to occupy the positions in the Ga$X_4$ tetrahedron, while S prefers to occupy the positions in the V$_4X_4$ heterocubane-like cluster.
The solid solutions do not undergo  a structural phase transition and  do not exhibit long-range ferromagnetic magnetic ordering. However an irreversible bifurcation between ZFCW and FCW traces in the $\chi$ vs $T$ measurements is observed in these materials, that
 probably reflects the presence of a spin glass-like state.
From the point of view of our interpretation, this implies that substitution of cations on either the $B$ or $A$ site has a profound effect on the ability of the material to undergo a structural phase transition.
However, there is still an appreciable electronic response in this temperature region, demonstrated by the 
 drastic decrease of $\mu_\text{eff}$, observed in the $\mu_\text{eff}$ vs $T$ curves depicted in Fig.~\ref{Fig:mag_mes}c.

The mechanism preventing a structural phase transition in intermediate compositions ($y = 1$ to 7)  requires further investigation.
However, in general terms we can comment that
the orbital degeneracy of V$^{3+}$-based compounds would often be expected to be
 lifted by a Jahn-Teller (JT) distortion which, if it happens at each
 site via a coherent lattice distortion, implies orbital
 ordering \cite{mazin2007charge}. Generally, three energy scales decide whether
 this occurs. These are the on-site Coulomb repulsion $U$, the electronic
 bandwidth $W$ and intra-atomic Hund rule coupling $J_{\mathrm
   {h}}$. In an insulator, $U\gg W$ and we expect a JT
 distortion; in a metal (where $W\gg U$), the JT distortion is
 suppressed.
Where $U$ and $W$ are similar, it is also possible \cite{mazin2007charge} that the Hund's rule
coupling overcomes on-site repulsion and leads to a charge ordering,
involving double electron occupancy on some sites
(V$^{3+}$) and single occupancy on others (V$^{4+}$). This charge ordering occurs via a mechanism involving a delocalization effect, where some of the electronic energy levels become band-like and consequently reduce the energy penalty for the double electron occupancy required for charge ordering \cite{mazin2007charge}. 
In the compounds GaV$_{4}$S$_{8}$ and GaV$_{4}$Se$_{8}$,
we might expect that their insulating nature puts them in the JT
regime. However, the possible delocalization of charges
over V$_{4}$ clusters that occurs in these systems, could push the
systems into the regime where $U$ and $W$ are similar and explain how
the charge ordering mechanism leads to the distortion in these materials. 

At a JT transition \cite{yosida} the distortion of the structure costs elastic
energy as $\alpha q^{2}$ (where $q$ is a displacement coordinate and $\alpha$ a
constant) while the resulting splitting of the electronic states
(through local crystal field levels and Hund's rule coupling) saves
energy as $-\beta q$ (where $\beta$ is a constant). An equilibrium distortion
(with $q_{0}=\beta/2 \alpha$) leads to a
net saving in energy and a JT distortion results.
To avoid a JT distortion in the intermediate materials, we could
imagine either
(i) the distortion costs too much energy or (ii) the electronic energy saving is not large enough, or both. 
In the charge-ordering scenario, (i) still applies in cases where there is a
distortion,
but here the electronic energy saving occurs because delocalization leads to a reduction
in the Coulomb energy cost of having two electrons on a site. In this case, 
any reduction in the ability of electrons to diffuse between sites can
reduce delocalization and hence also suppress the transition.  

Since the end members of the series have similar electronic
behaviour,
it seems unlikely that S/Se disorder radically alters the
electronic structure, such that these materials lie in a different
energetic regime.
Indeed, the behavior of  $\mu_{\mathrm{eff}}$ suggests that all of the
compounds are susceptible to a change in the electronic structure around
40~K. (However, we note that the $\mu^{+}$SR results suggested that the ground
states of the two end systems are quite different to each other, so a change in
the electronic states with substitution cannot be ruled out.)
To explain the lack of structural transition in the intermediate
materials, 
it is more probable that either the S/Se disorder creates a large
elastic energy cost for a structural distortion or, if the distortion
is driven by charge ordering, that electron delocalization is
suppressed by the S/Se disorder. 
In the first hypothesis, the uneven local distribution of S and Se in the  V$_4X_4$ units, where neighbouring units contain slightly different concentrations of S and Se, could prevent a structural distortion and therefore retain the material in the cubic symmetry.
The second hypothesis involves the disorder  shutting down electron diffusion pathways via S and Se anions, reducing electron delocalization and consequently preventing the mechanism for charge ordering of V cations, which seems to be the driving force for the structural phase transition.
Finally we note that while our proposed charge-ordering scenario provides a plausible explanation for the magnetic properties discussed in this paper, whether it also provides a straightforward and complete description of other (non-magnetic) behaviour of this system remains an open question, deserving of future investigation.

All of the results of our investigation on this series, when taken together, provide a direct link between the structural phase transitions and the observation of magnetic transitions in the materials. This is the case in the two end members, \ce{GaV4S8} and \ce{GaV4Se8}, which exhibit both these features whilst the intermediate compositions investigated exhibit neither of these transitions. The two parent compounds still remain the only members which host a skyrmion phase, making them unique, and the observation of both of the transitions appear to be prerequisites for the stabilization of the skyrmion phases. 

Polycrystalline and single crystal samples of the entire series of compounds GaV$_4$S$_{8-y}$Se$_y$ $( y= 0$ to $8)$ have been synthesized and single crystals of the materials obtained by the CVT technique.
Both the polycrystalline and single crystal samples have been thoroughly investigated for the structural properties through the use of powder and single crystal x-ray as well as neutron diffraction.
Additionally, the magnetic properties of the whole series of materials have been studied using \textit{dc} and \textit{ac} magnetic susceptibility, magnetization and heat capacity measurements.
Our results show that while both GaV$_{4}$S$_{8}$ and GaV$_{4}$Se$_{8}$ both undergo structural phase transitions, followed by magnetic ordering at 13 and 18~K respectively, all the intermediate compositions retain their cubic symmetry down to the lowest temperature investigated (1.5~K).
These intermediate compositions do not exhibit long-range magnetic order but are, however, found to exhibit a spin glass-like behaviour with freezing temperatures of about 2.5~K, as ascertained from \textit{ac} susceptibility and heat capacity measurements.
Examining the magnetic structure of GaV$_{4}$S$_{8}$ in detail, there is a clear indication of the localization of spins on the V atoms within the V$_4X_4$ cluster which appears to favour a charge ordering mechanism.
The results presented form a thorough investigation of both the structural and magnetic properties of this entire family of interesting materials.
In the search for additional members of this family beyond the two end compositions to host skyrmions, this study throws light on the crucial links that exist between the structure and magnetism in these materials.
Given the subtle nature of the local structural distortions in the intermediate compositions as a function of temperature, one strategy that emerges from our study is the possibility that investigating materials with much smaller substitutional levels close to the two end members ($y$ between 0.1 and 0.5, and 7.5 and 7.9) maybe promising.
The study of these materials forms part of our ongoing investigations to provide additional insight into structure-property correlations in this interesting family.

\section*{Methods}

\subsection*{Synthesis and crystal growth}
Polycrystalline GaV$_4$S$_{8-y}$Se$_{y}$ materials with $y$ between 0 and 8 were synthesized by thoroughly grinding stoichiometric amounts of V ($99.5\%$, metals basis, Sigma-Aldrich), S ($99.99\%$, Acros Organics), and Se ($99.999\%$, Alfa Aesar) powders inside an argon-filled glove box.
The mixtures of powders were transferred into silica tubes along with stoichiometric amounts of Ga ($99.999\%$, metals basis, Alfa Aesar) granules and 50~mg of \ce{I2} ($99.99\%$, Alfa Aesar; added to prevent the formation of \ce{V5S8}).
The tubes were then evacuated and sealed.
The mixtures were heated at a rate of $10~^\text{o}$C/h to $600~^\text{o}$C and at a rate of 5$~^\text{o}$C/h to 820$~^\text{o}$C, kept at this temperature for 200~h, followed by water quench cooling.
This yielded black polycrystalline air-stable powders with metallic lustre.

Single crystals of GaV$_4$S$_{8-y}$Se$_{y}$ were grown using the chemical vapour phase transport (CVT) technique.
Polycrystalline materials (1.5~g), described above, and ~3.3 mg/cc of the transporting agent \ce{I2} or \ce{PtCl2}, were sealed in evacuated silica tubes.
The growth of single crystals was achieved by heating the source part of the tube to 900$~^\text{o}$C and the sink part to 850$~^\text{o}$C for four weeks.
Black octahedral single crystals with well-defined facets of various sizes ranging from several microns up to $2 \times 2 \times 2$ mm$^3$ (Supplementary Fig.~S1) were obtained.

\subsection*{Energy Dispersive X-ray Measurements}
Energy dispersive X-ray (EDX) analysis was performed on single crystals attached to conductive carbon tabs with a facet parallel to the sample stub, using a scanning electron microscope (Zeiss Gemini500) equipped with an XMAX150 Detector (Oxford Instruments Analytical system).
The 25~keV energy beam and K lines were used to determine the atomic percentage of Ga, V, S, and Se atoms in the crystals.

\subsection*{Powder and Single Crystal X-ray Diffraction}
The phase purity determination and structural investigations of the polycrystalline materials were carried out using a Panalytical X-Pert Pro diffractometer operating in Bragg-Brentano geometry, equipped with monochromatic Cu K$_{\alpha 1}$ $(\lambda = 1.5406$~\AA) source and solid-state PIXcel one-dimensional detector in the $2\Theta$ range between 10 and 90$^\text{o}$ (2~h scan) at ambient temperature.
Rietveld refinements were carried out using the TOPAS academic v6.0 software \cite{Coelho2018TOPAS}.

Single crystal X-ray diffraction data were collected on $\sim 50 - 100$~$\mu$m sized crystals at 120 and 30~K using a Rigaku Oxford diffraction SuperNova diffractometer, equipped with dual wavelength (Cu/Mo) microfocus X-ray source (Mo K$_{\alpha}$, $\lambda = 0.71073$~\AA), Atlas S2 CCD area detector and Oxford Cryosystems N-Helix cryo cooling system.
The structures were solved by the direct method and refined by full-matrix least squares on $F^2$ for all data using SHELXT \cite{sheldrick2015shelxt} and SHELXL \cite{sheldrick2015crystal} embedded in the Olex2 \cite{dolomanov2009olex2} software.

\subsection*{Powder and Single Crystal Neutron Diffraction}
Two complementary sets of powder neutron diffraction experiments were performed at the Institute Laue Langevin (ILL) on the D2B and D20 diffractometers, to determine the nuclear and magnetic structures of two of the materials in the series, namely \ce{GaV4S8} and \ce{GaV4S6Se2}.
The D2B diffractometer was used to obtain high-resolution diffractograms for nuclear structure characterization using a neutron wavelength of $1.594$ \AA \ over the temperature range $1.5-50$~K.
High-intensity powder neutron diffraction was performed on the D20 diffractometer using a wavelength of $2.417$~\AA \ to determine the magnetic scattering over the temperature range $1.5-50$~K.
The \ce{GaV4S8} and \ce{GaV4S6Se2} powder samples ranging from 3-4 g were sealed in thin-walled cylindrical vanadium containers of 8~mm diameter and placed in a standard orange cryostat.
Structural and magnetic Rietveld refinements were carried out using the TOPAS academic v6.0 and FullProf software \cite{Rodriguez1993recent}.

Single crystal neutron diffraction experiments were preformed using the SXD diffractometer at the ISIS Neutron and Muon spallation source \cite{keen2006sxd}. 
Data was collected on $\sim 2\times 2 \times 2$~mm$^3$ crystals of \ce{GaV4S4Se4} and \ce{GaV4Se8}. 
The samples were mounted in an aluminum loop on top of an Al pin and were placed in a vanadium cryostat which was evacuated to $\sim 10^{-5}$~mbar.
Data were processed using the SXD2001 software and intensities extracted using a least squares procedure \cite{keen2006sxd} with 3D Gauss-ellipsoids, line fittings, and box fittings with time-of-flight asymmetry as the profile function.
The nuclear structures were refined using the JANA software \cite{petvrivcek2014crystallographic}.

\subsection*{Magnetic Measurements}
Quantum Design Magnetic Property Measurement System, MPMS-5S and MPMS-7XL, superconducting quantum interference device (SQuID) magnetometers were used for the investigation of the magnetic properties of the polycrystalline samples and single crystals as a function of temperature and field.
\textit{dc} temperature-dependent magnetic susceptibility $(\chi)$ measurements were carried out in an applied field of 10~mT and 2~T in the temperature region between 1.8 and 300~K under both zero-field-cooled-warming (ZFCW) and field-cooled-warming (FCW) protocols.
The magnetic moments were obtained from $M$ vs\ \textit{B} measurements in applied fields between 0 and 5~T at 1.8~K.
\textit{ac} frequency-dependent susceptibility ($\chi'_\text{ac}$ and $\chi''_\text{ac}$) measurements were performed on non-oriented single crystals at zero applied magnetic field between 1.8 and 3.5~K.

Heat capacity measurements were performed in a Quantum Design Physical Property Measurement System from $1.8\leq T \leq 300$~K.
A He$^3$ insert was used to obtain temperatures down to  0.5~K.

\section*{Acknowledgements}

This work is financially supported by EPSRC (EP/N032128/1). Part of this work was performed at the Science and Technology Facilities Council (STFC) ISIS Facility and the Institut Laue-Langevin (ILL). We are grateful for the provision of beamtime. The authors would like to acknowledge ILL for beam time allocation under the experiment codes 5-31-2608. Data are available from ILL at DOI: 10.5291/ILL-DATA.5-31-2608. We would like to thank Matic Lozin\v{s}ek (Univeristy of Ljubljana and Jo\v{z}ef Stefan Insitute) for help with solving crystal structures and fruitful discussions. We also thank Steve York (University of Warwick) for EDX analysis and David Walker (University of Warwick) for assistance with single crystal X-ray diffraction measurements.  Crystallographic data were collected using an instrument that received funding from the ERC under the European Union’s Horizon 2020 research and innovation programme (Grant Agreement No. 637313).



\newpage
\section*{\LARGE Supplementary Information}
\renewcommand\thefigure{S\arabic{figure}} 
\setcounter{figure}{0}
\renewcommand\thetable{S\arabic{table}} 
\setcounter{table}{0}
\begin{table}[htb]
\centering
\caption{Refined parameters of GaV$_4$S$_{8-y}$Se$_{y}$ $(y = 0$ to $8)$ in the $F\bar{4}3m$ space group at room temperature from powder X-ray diffraction data taken on a laboratory diffractometer.}
\label{tab:PXRD}
\small
\resizebox{\textwidth}{!}{
    \begin{tabular}{cccccccccc}
        \hline
        $y$  &0&1&2&3&4&5&6&7&8  \\
        \hline
        $a$~\AA & 9.66294(7)& 9.71632(6)&9.78401(4)&9.85542(9)&9.92004(6)&9.98498(5)&\makecell{9.9800(2) \\ 10.03670(16)}&10.09205(7)&10.14102(3) \\
        $V~($\AA$^3)$ & 902.25(2)&	917.288(16)&	936.593(12)&	957.25(2)&	976.203(17)&	995.500(16)&	\makecell{994.02(7)\\ 1011.05(5)}&	1027.87(2)&	1042.905(8)\\
        sof S1&	1	&0.910(10)&	0.719(11)&	0.574(8)&	0.413(9)&	0.176(10)&\makecell{	0.06(4) \\ 0.096(18)}	&0.05&	N/A\\
        sof S2& 	1&	0.938(11)&	0.882(10)&	0.798(9)&	0.692(8)&	0.497(10)&	\makecell{0.55(5)\\ 0.42(2)} &	0.230(14)&	N/A\\
        sof Se1&	N/A &	0.090(10)&	0.281(11)&	0.426(8)&	0.587(9)&	0.824(10)&	\makecell{0.94(4)\\ 0.904(18)}&	0.95&	1\\
        sof Se2&	N/A	&0.062(11)&	0.118(10)&	0.202(9)&	0.308(8)&	0.503(10)&	\makecell{0.45(5)\\ 0.58(2)}&	0.770(14)&	1\\
        Formula based on sof&	\ce{GaV4S8}&	GaV$_{4}$S$_{7.39}$Se$_{0.61}$&	GaV$_{4}$S$_{6.40}$Se$_{1.60}$&	GaV$_{4}$S$_{5.49}$Se$_{2.51}$&	GaV$_{4}$S$_{4.42}$Se$_{3.58}$&	GaV4S$_{2.69}$Se$_{5.31}$&\makecell{GaV4S$_{2.44}$Se$_{5.56}$ \\ 23.3(9)\%\\GaV$_{4}$S$_{2.06}$Se$_{5.94}$ \\76.7(9)\%} &	GaV$_{4}$S$_{1.12}$Se$_{6.88}$&	GaV$_{4}$Se$_{8}$\\
        Formula weight (g/mol)&	530.01&	558.62&	605.04&	647.72&	697.89&	779.02&\makecell{790.75\\ 808.75}&847.49&	905.17\\
        Phase purity&	Phase pure&	Phase pure&	Phase pure&	Phase pure&	Phase pure&	Phase pure&	Two phases&	\ce{VSe2} impurity 3.3\% &	\ce{VSe2} impurity 2.0\% \\
        $R_{wp}$&	17.07&	16.95&	16.99&	18.08&	17.65&	17.77&	26.81&	31.10\textsuperscript{*}&	23.21\textsuperscript{*}\\
        $R_{exp}$&	14.90&	13.90&	13.32&	14.12&	13.47&	12.48&	19.40&	9.47&	12.50\\
        $\chi^2$&	1.31&	1.49&	1.63&	1.64&	1.72&	2.03&	1.91&	10.79\textsuperscript{*}&	3.45\textsuperscript{*}\\
        Atomic positions\\
        Ga on 4a &0&0&0&0&0&0&\makecell{0\\0}&0&0\\
        V on 16e (x, x, x)& 0.60508(14) & 0.60516(14) & 0.60487(14) & 0.60442(16) & 0.60336(18) & 0.60327(18) & \makecell{0.6042(8) \\ 0.6034(3)} & 0.6049(4) & 0.6001(3)\\
        S1 on 16e (x, x, x)&0.3709(2)&	0.3730(7)&	0.3722(7)&	0.369(4)&	0.368(6)&	0.3790(15)& \makecell{0.37(3)\\ 0.374(6)}&	0.398(2)&	N/A\\
        S2 on 16e (x, x, x)&	0.8631(2)&	0.8634(2)&	0.86460(19)&	0.86454(19)&	0.86399(18)&	0.86237(16)&\makecell{0.8656(6)\\ 0.8649(3)}&	0.8637(3)&	N/A\\
        Se1 on 16e (x, x, x)&	N/A&	0.359(4)&	0.360(2)&	0.367(7)&	0.369(6)&	0.3671(5)&	\makecell{0.366(16)\\ 0.3671(18)}&	0.3685(4)&	0.36869(18)\\
        Se2 on 16e (x, x, x)&	N/A&	0.8634(2)&	0.86460(19)&	0.86454(19)&	0.86399(18)&	0.86237(16)&\makecell{0.8656(6)\\ 0.8649(3)}&	0.8637(3)&	0.85940(18)\\
        
        \hline
    \end{tabular}
    }
    \textsuperscript{*} Higher R$_\text{wp}$ and $\chi^2$ values are due to the presence of the preferred orientation in \ce{VSe2} (trigonal, $P\bar{3}m1$).
\end{table}

\begin{table}
\centering
\caption{Crystallographic data of GaV$_4$S$_{8-y}$Se$_y$ $(y = 0$ to $8)$ single crystals obtained from single crystal X-ray diffraction performed on a laboratory diffractometer.}
\label{tab:CXRD_1}
\resizebox{\textwidth}{!}{
    \begin{tabular}{cccccccccc}
        \hline
        $y$  &0&1&2&3&4&5&6&7&8  \\
        \hline
        a (\AA)&9.64873(6)&9.67159(11)&9.73879(9)&9.80152(13)&9.88086(12)&9.96041(15)&10.01498(9)&10.06360(10)&10.12510(10)\\
        V (\AA$^3$)&898.277(18)&904.68(3)&923.67(3)&941.63(4)&964.68(4)&988.17(5)&1004.50(3)&1019.20(3)&1038.00(3)\\
        Z&4&4&4&4&4&4&4&4&4\\
        sof S1&1&0.969(3)&0.904(3)&0.822(4)&0.713(5)&0.553(7)&0.407(8)&0.255(8)&N/A\\
        sof S2 &1&0.934(4)&0.783(3)&0.599(4)&0.377(6)&0.237(8)&0.143(8)&0.047(8)&N/A\\
        sof Se1&N/A&0.031(3)&0.097(3)&0.178(4)&0.287(5)&0.447(7)&0.593(8)&0.745(8)&1\\
        sof Se2&N/A&0.066(4)&0.217(3)&0.401(4)&0.623(6)&0.763(8)&0.857(8)&0.953(8)&1\\
        Empirical formula&S$_8$&S$_{7.61}$Se$_{0.39}$&S$_{6.74}$Se$_{1.25}$&S$_{5.68}$Se$_{2.32}$& S$_{4.20}$Se$_{3.80}$&S$_{3.12}$Se$_{4.88}$  &S$_{2.20}$Se$_{5.80}$  &S$_{1.09}$Se$_{6.91}$&Se$_8$ \\
        &&&&&&&&&\\
        M$_\text{R}$ (g/mol)&529.6&548.13&588.82&638.53&700.62&756.6&801.92&849.11&905.09\\
        $\rho_\text{calc}$ (g/cm$^3$)&3.916&4.024&4.234&4.504&4.824&5.086&5.303&5.534&5.792\\
        Crystal size (mm3)&0.137 $\times$ 0.116 $\times$ 0.09&0.075 $\times$ 0.074 $\times$ 0.052&0.087 $\times$ 0.064 $\times$ 0.061&0.105 $\times$ 0.087 $\times$ 0.059&0.146 $\times$ 0.126 $\times$ 0.098&0.078 $\times$ 0.07 $\times$ 0.069&0.134 $\times$ 0.117 $\times$ 0.078&0.148 $\times$ 0.101 $\times$ 0.097&0.124 $\times$ 0.101 $\times$ 0.055\\
        $\mu$/mm$^{-1}$&8.75&10.161&13.174&16.785&21.097&24.736&27.639&30.853&34.019\\
        Absor. correction&Analytical&Analytical&Analytical&Analytical&Analytical&Analytical&Analytical&Analytical&Analytical\\
        $T_\text{min}$&0.445&0.472&0.429&0.308&0.131&0.231&0.103&0.066&0.097\\
        $T_\text{max}$&0.558&0.576&0.517&0.427&0.264&0.314&0.247&0.192&0.271\\
        $2\theta$ range $\left(^{\circ}\right)$&7.314 to 89.284&7.298 to 90.1&7.248 to 89.464&7.2 to 89.794&7.142 to 90.066&7.086 to 89.882&7.048 to 89.838&7.014 to 89.86&6.97 to 89.732\\
        reflection collected&17161&1596&2872&2447&3336&2380&6250&16197&5452\\
        indep. Reflect.&420&420&430&438&441&454&470&472&481\\
        restrains/ paramet.&0/12; 0/13&0/16&0/17&0/17; 0/18&0/17&0/18&0/17&0/17&0/12; 0/13\\
        $R_\text{int}$&0.0493&0.0191&0.0151&0.0151&0.0189&0.0182&0.0358&0.0603&0.0343\\
        $R_1$, $wR_2$ $\left[I\geq 2\delta~(I)\right]$ &0.0118, 0.0293&0.0143, 0.0326&0.0100, 0.0249&0.0120, 0.0298&0.0115, 0.0318&0.0154, 0.0387&0.0146, 0.0387&0.0142, 0.0361 &0.0143, 0.0369\\
        Goodness-of-fit on $F^2$ &1.251&1.069&1.137&1.169&1.262&1.158&1.143&1.192&1.2\\
        Largest diff. peak/hole $\left(\text{e} \text{\AA}^{-3}\right)$&0.50/ -0.68&0.34/ -0.45&0.30/ -0.44&0.37/ -0.57&0.54/ -0.74&0.47/ -0.67&0.62/ -0.42&0.53/ -0.59&1.17/ -0.92\\
        Flack paramet.&0.010(10); N/A&-0.021(18)&0.012(9)&-0.011(15); N/A&0.011(14)&N/A&-0.011(15)&0.007(15)&0.003(10); N/A\\
        twin ratio&N/A; 0.984(12)&N/A&N/A&N/A; 1.006(19)&N/A&0.14(2)&N/A&N/A&N/A; 0.993(19)\\
        \hline
    \end{tabular}
    }
\end{table}

\begin{table}
\centering
\caption{Crystallographic data for GaV$_4$S$_{8-y}$Se$_y$ $(y = 0$ to $8)$ single crystals along with selected bond distances, angles and tetrahedrons volumes from single crystal X-ray diffraction performed on a laboratory diffractometer.
Bond distances, angles, and tetrahedral volumes are given in formats X-Y, X-Y-Z, and \ce{X4} tetrahedra, respectively.}
\label{tab:CXRD_2}
\resizebox{\textwidth}{!}{
    \begin{tabular}{p{3.5cm}p{2.2cm}p{2.2cm}p{2.2cm}p{2.2cm}p{2.2cm}p{2.2cm}p{2.2cm}p{2.2cm}p{2.2cm}}
        \hline
        $y$  &0&1&2&3&4&5&6&7&8  \\
        \hline
        \\
        \multicolumn{10}{c}{Atomic positions}\\
        \\
        Ga on 4a&0&0&0&0&0&0&0&0&0\\
        V on 16e (x, x, x)&0.39407(2); 0.60593(2)&0.60576(2)&0.60529(2)&0.39527(2); 0.60473(2)&0.60408(2)&0.60393(3)&0.60389(3)&0.60397(4)&0.39553(3); 0.60447(3)\\
        S1 on 16e (x, x, x)&0.62949(3); 0.37051(3) &0.37072(18)&0.37166(17)&0.6279(3); 0.3721(3)&0.3737(3)&0.3745(7)&0.3751(7)&0.3756(13)&N/A\\
        S2 on 16e (x, x, x)&0.13590(3); 0.86410(3) &0.86418(4)&0.86426(2) &0.13584(2); 0.86416(2)&0.86387(2)&0.86367(3)&0.86349(3)&0.86336(2)&N/A\\
        Se1 on 16e (x, x, x)&N/A&0.364(2)&0.3629(6)&0.6354(5); 0.3646(5)&0.3647(3)&0.3659(3)&0.36664(19)&0.36741(18)&0.63174(2); 0.36826(2) \\
        Se2 on 16e (x, x, x)&N/A&0.86418(4)&0.86426(2)&0.13584(2); 0.86416(2)&0.86387(2)&0.86367(3)&0.86349(3)&0.86336(2)&0.13668(2); 0.86332(2)\\
        \\
        \multicolumn{10}{c}{Selected bond distances in \AA, angles in $^\circ$ and tetrahedral volumes in \AA$^3$}\\
        \\
        V - S1 &2.2941(4)&2.2959(18)&2.2973(17)&2.303(3)&2.297(3)&2.305(7)&2.311(7)&2.317(13)&N/A\\
        V - Se1&N/A&2.375(19)&2.401(6)&2.392(5)&2.405(3)&2.409(3)&2.4124 (19)&2.4152(19)&2.4233(4)\\
        V-V (intra)&2.8909(4)&2.8931(4)&2.9003(4)&2.9034(3)&2.9088(3)&2.9279 (6)&2.9429(6)&2.9594(7)&2.9918(5)\\
        V-S1-V&78.11(1)&78.11(6)&78.28(5)&78.17(9)&78.55(9)&78.8(2)&79.1(2)&79.4(4)&N/A\\
        V-Se1-V&N/A&75.1(6)&74.3(2)&74.74(14)&74.41(9)&74.86(9)&75.17(6)&75.56(6)&76.24(1)\\
        S1-V-Se1&N/A&2.0(5)&2.57(15)&2.22(14)&2.68(10)&2.59(19)&2.56(18)&2.5(3)&N/A\\
        V-V (inter)&3.9318(3)&3.9457(4)&3.9861(4)&4.0276(5)&4.0786(5)&4.1151(6)&4.1388(6)&4.1566 (7)&4.1677(6)\\
        V$_4$ tetrahedra&2.8473(9)&2.8538(9)&2.8750(10)&2.8844(10)&2.9004(10)&2.9582(15)&3.0036(15)&3.055(2)&3.1560(16)\\
        S1$_4$ tetrahedra&5.201(2)&5.213(13)&5.207(12)&5.25(2)&5.18(2)&5.21(5)&5.22(5)&5.23(10)&N/A\\
        Se1$_4$ tetrahedra&N/A&6.07(15)&6.35(5)&6.23(4)&6.37(2)&6.35(3)&6.353(16)&6.335(15)&6.3288(17)\\
        V$_4$cent-V$_4$cent&6.82268(3)&6.83885(6)&6.88636(5)&6.93072(7)&6.98682(6)&7.04307(8)&7.08166(5)&7.11604(5)&7.15953(5)\\
        V-S2/Se2&2.5244(4)&2.5329(4)&2.5567(3)&2.5791(3)&2.6057(3)&2.6271(4)&2.6406(4)&2.6515(5)&2.6612(4)\\
        Ga-S2/Se2&2.2712(3)&2.2752(4)&2.2897(2)&2.3061(2)&2.3297(2)&2.3520(3)&2.3680(3)&2.3817(2)&2.3970(2)\\
        Ga-S2/Se2-V&115.941(13)&115.923(16)&115.823(9)&115.639 (9)&115.368 (9)&115.260(13)&115.188 (13)&115.165 (12)&115.284(10)\\
        Ga-V&4.0678(2)&4.0781(2)&4.1084(2)&4.1373(2)&4.1736(2)&4.2079(3)&4.2311(3)&4.2513(4)&4.2750(3)\\
        Ga-Ga&6.82268(3)&6.83885(6)&6.88636(5)&6.93072(7)&6.98682(6)&7.04307(8)&7.08166(5)&7.11604(5)&7.15953(5)\\
        S2/Se2$_4$ tetrahedra&6.012(2)&6.044(3)&6.1604(16)&6.2941(16)&6.4895 (17)&6.677(3)&6.814(3)&6.9337(18)&7.0678(18)\\
    
        \hline
    \end{tabular}
    }
\end{table}

\begin{table}
\centering
\caption{Powder neutron diffraction refined parameters of GaV$_4$S$_{8}$ on D2B diffractometer at the ILL.}
\label{tab:PND_D2B}
\small
\resizebox{\textwidth}{!}{
    \begin{tabular}{ccccccc}
        \hline
        Temperature (K) & 50&&30&15&9&1.5  \\
        \hline
        Space Group & $F\bar{4}3m$& & $R3m$ & $R3m$& $R3m$& $R3m$\\
        $Z$ & 4& & 3 &3 &3 &3\\
        $a$~\AA  &9.64823(5)&& 6.80183(6)& 6.80101(6) & 6.80098(6)& 6.80084(6)\\
        $c$~\AA &  N/A& &16.7984(3)& 16.8034(3)& 16.8058(3)& 16.8053(3)\\
        $V~($\AA$^3)$ & 898.139(13)& & 673.055(18)& 673.091(18) & 673.182(17)& 673.133(18)\\
        $R_{wp}$  & 4.79& & 4.61 &4.62& 4.52& 4.53\\
        $R_{exp}$  & 3.77& & 3.77 & 3.77& 3.77& 3.77\\
        $\chi^2$ &  1.62& & 1.49 & 1.50 & 1.44 & 1.44\\
        \\
        \multicolumn{7}{c}{Atomic positions}\\
        \\
        Ga on 4a&  0& Ga on 3a (z fixed) & 0 & 0 & 0& 0\\
        S1 on 16e (x, x, x) & 0.37023(12)&S1 on 3a (0, 0, z) & 0.6312(5)&0.6314(4)& 0.6308(4)& 0.6312(4)\\
        S2 on 16e (x, x, x) &0.86405(11)& S2 on 9b (x, 2x, z) & 0.1719(3) 0.4551(3)& 0.1716(3) 0.4549(3)&0.1716(3) 0.4547(3)&0.1716(3) 0.4548(3)\\
        &&S3 on 9b (x, 2x, z) &0.1802(3) 0.9525(3) & 0.1796(3) 0.9520(3)&0.1797(3) 0.9522(3)&0.1800(3) 0.9523 (3)\\
        &&S4 on 3a (0, 0, z) & 0.1347(4)&0.1353(4)& 0.1357(4)& 0.1356(4)\\
        \hline
    \end{tabular}
    }
\end{table}

\begin{table}
\centering
\caption{Single crystal neutron diffraction refined parameters of GaV$_4$S$_4$Se$_{4}$ from data collected on the SXD diffractometer at ISIS. Positions of the V atoms are constrained to those from the single crystal X-ray diffraction.}
\label{tab:SXD_GaV4S4Se4}
    \begin{tabular}{ccc}
        \hline
        Temperature (K)  &55&1.5 \\
        \hline
        Space Group & $F\bar{4}3m$ & $F\bar{4}3m$ \\
        $Z$ & 4 & 4\\
        $a$~\AA &9.89110(10)& 9.89250(10) \\
        $V~($\AA$^3)$ & 967.684(17) & 968.095(17)\\
        M$_\text{R}$ (g/mol)& 700.7 & 700.7\\
        $\rho_\text{calc}$ (g/cm$^3$)& 4.8094 & 4.8074 \\
        Crystal size (mm$^3$)& $1 \times\ 1.5  \times\ 1.5$ & $1 \times\ 1.5  \times\ 1.5$ \\
        $2\theta$ range $\left(^{\circ}\right)$ &3.61 to 73.22& 3.56 to 73.25\\
        reflection collected&512 & 1798\\
        indep. Reflect.& 509 & 1784\\
        constraints/ paramet.& 5/12 & 5/12\\
        $R_\text{int}$& & \\
        $R_1$, $wR_2$ $\left[I\geq 2\delta~(I)\right]$ & 0.0804, 0.1610 & 0.0804, 0.1506\\
        Goodness-of-fit on $F^2$ & 2.03 & 2.65\\
        \\
        \multicolumn{3}{c}{Atomic positions}\\
        \\
        Ga on 4a & 0 & 0 \\
        V on 16e (x, x, x) & 0.60408 & 0.60408 \\
        S1 on 16e (x, x, x) & 0.8638(2) &0.86401(8)\\
        Se1 on 16e (x, x, x) & 0.8638(2) & 0.86401(8)\\
        S2 on 16e (x, x, x) & 0.86332(5) & 0.3717(5)\\
        Se2 on 16e (x, x, x) & 0.3740(11) & 0.3656(5)\\
        \hline
    \end{tabular}
\end{table}

\begin{table}
\centering
\caption{Single crystal neutron diffraction refined parameters of GaV$_4$Se$_{8}$ on SXD diffractometer at the ISIS.}
\label{tab:SXD}
    \begin{tabular}{cccc}
        \hline
        Temperature (K)  &55&&30 \\
        \hline
        Space Group & $F\bar{4}3m$ && $R3m$ \\
        $Z$ & 4 & & 3\\
        $a$~\AA &10.11750(10)& &7.1490(18) \\
        $c$~\AA & N/A & &17.524(6)\\
        $V~($\AA$^3)$ & 1035.666(18) & &775.6(4)\\
        M$_\text{R}$ (g/mol)& 905.2 & &905.2\\
        $\rho_\text{calc}$ (g/cm$^3$)& 5.8052 & &5.8135 \\
        Crystal size (mm$^3$)& 2 $\times$\  2 $\times$ 2&& 2 $\times$\ 2 $\times$ 2\\
        $2\theta$ range $\left(^{\circ}\right)$ &3.44 to 82.14&& 3.44 to 78.84\\
        reflection collected&4740 && 3637\\
        indep. Reflect.& 2948 && 2638\\
        constraints/ paramet.& 0/15 && 8/16\\
        $R_\text{int}$& && \\
        $R_1$, $wR_2$ $\left[I\geq 2\delta~(I)\right]$ & 0.0701, 0.1426 && 0.0652, 0.1276\\
        Goodness-of-fit on $F^2$ & 1.40 && 1.52\\
        \\
        \multicolumn{4}{c}{Atomic positions}\\
        \\
        Ga on 4a & 0 & Ga on 3a (z fixed)& 0 \\
        S1 on 16e (x, x, x) & 0.36822(5) &S1 on 3a (0, 0, z)& 0.63170(14)\\
        S2 on 16e (x, x, x) & 0.86332(5) &S2 on 9b (x, 2x, z)& 0.17595(9) 0.45560(12)\\
        & &S3 on 9b (x, 2x, z)&0.18184(10) 0.95444(12)\\
        & &S4 on 3a (0, 0, z)& 0.13636(14)\\
        \hline
    \end{tabular}
\end{table}

\begin{figure}[t]
\includegraphics[width=0.9\linewidth]{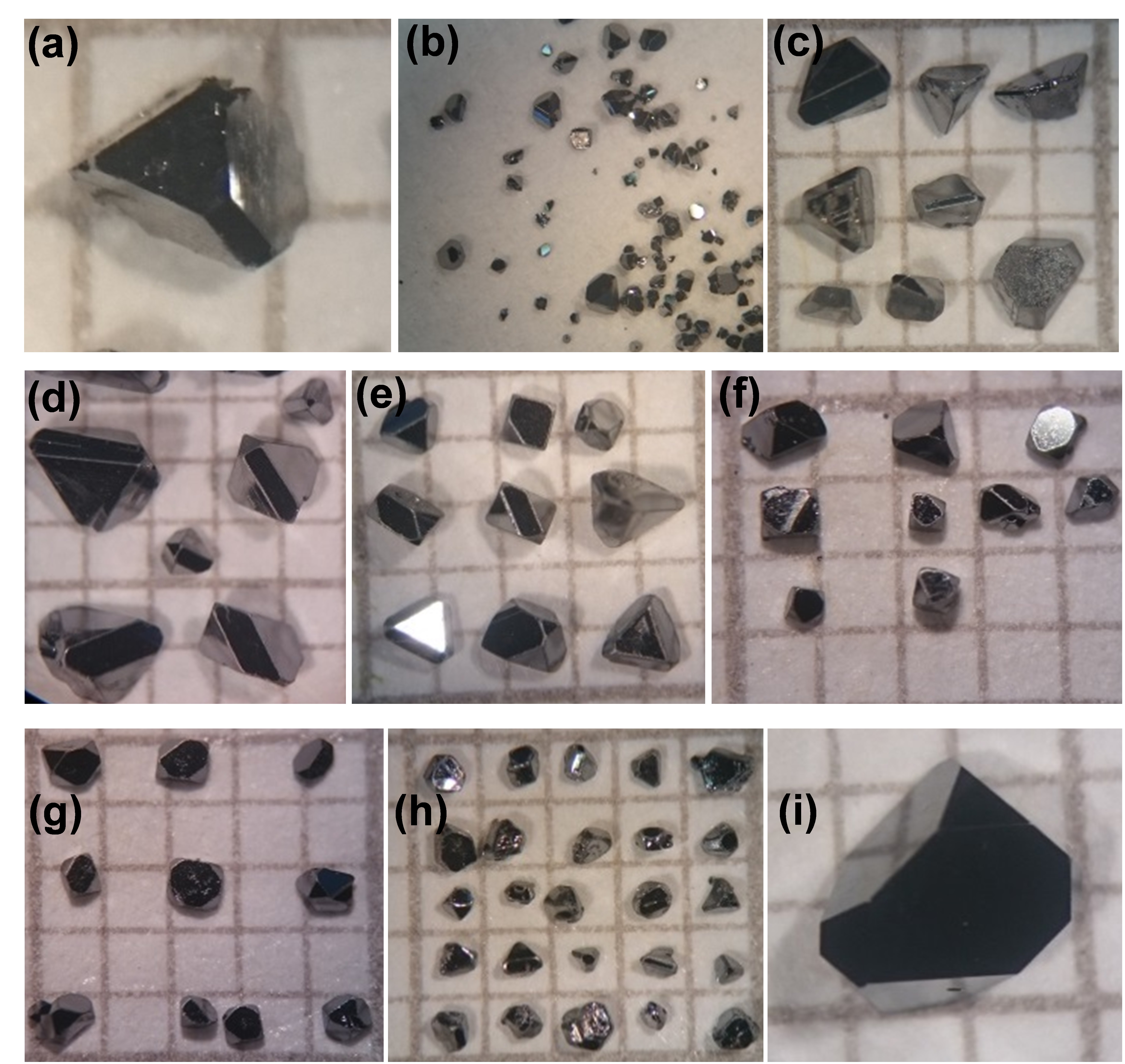}
\caption{Crystals of GaV$_4$S$_{8-y}$Se$y$ (a) $y=0$, (b) $y=1$, (c) $y=2$, (d) $y=3$, (e) $y=4$, (f) $y=5$, (g) $y=6$, (h) $y=7$, and (i) $y=8$ photographed on millimetre paper.}
\end{figure}

\begin{figure}[t]
\includegraphics[width=\linewidth]{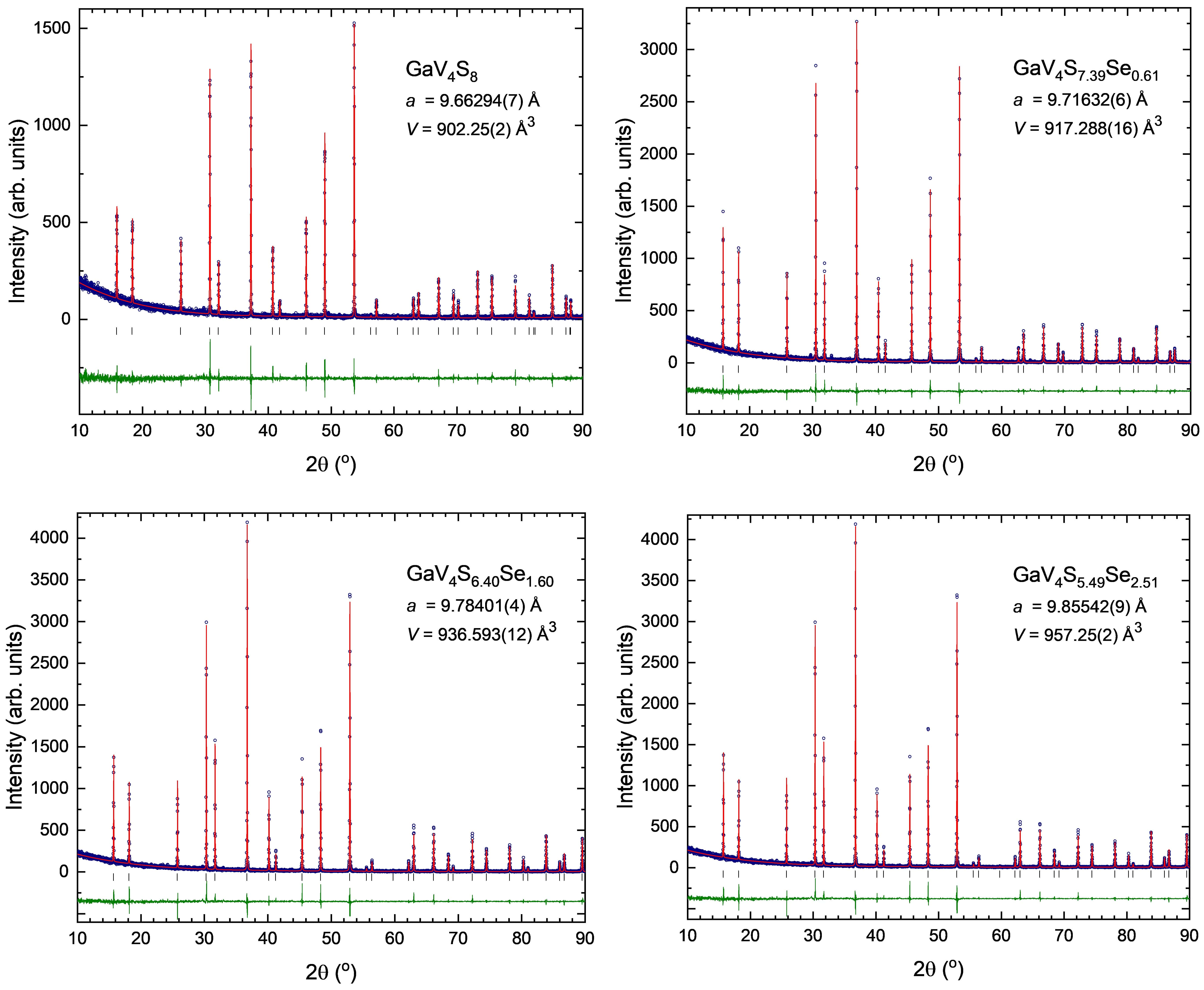}
\caption{Powder X-ray diffraction profiles of \ce{GaV4S8}, \ce{GaV4S7Se}, \ce{GaV4S6Se2}, and \ce{GaV4S5Se3}.
The experimentally-obtained diffraction profile at ambient temperature (blue open circles), refinement based on the model obtained from single crystal X-ray diffraction at room temperature (red solid line), difference (olive green solid line) and predicted peak positions (black tick marks).}
\end{figure}

\begin{figure}[t]
\includegraphics[width=\linewidth]{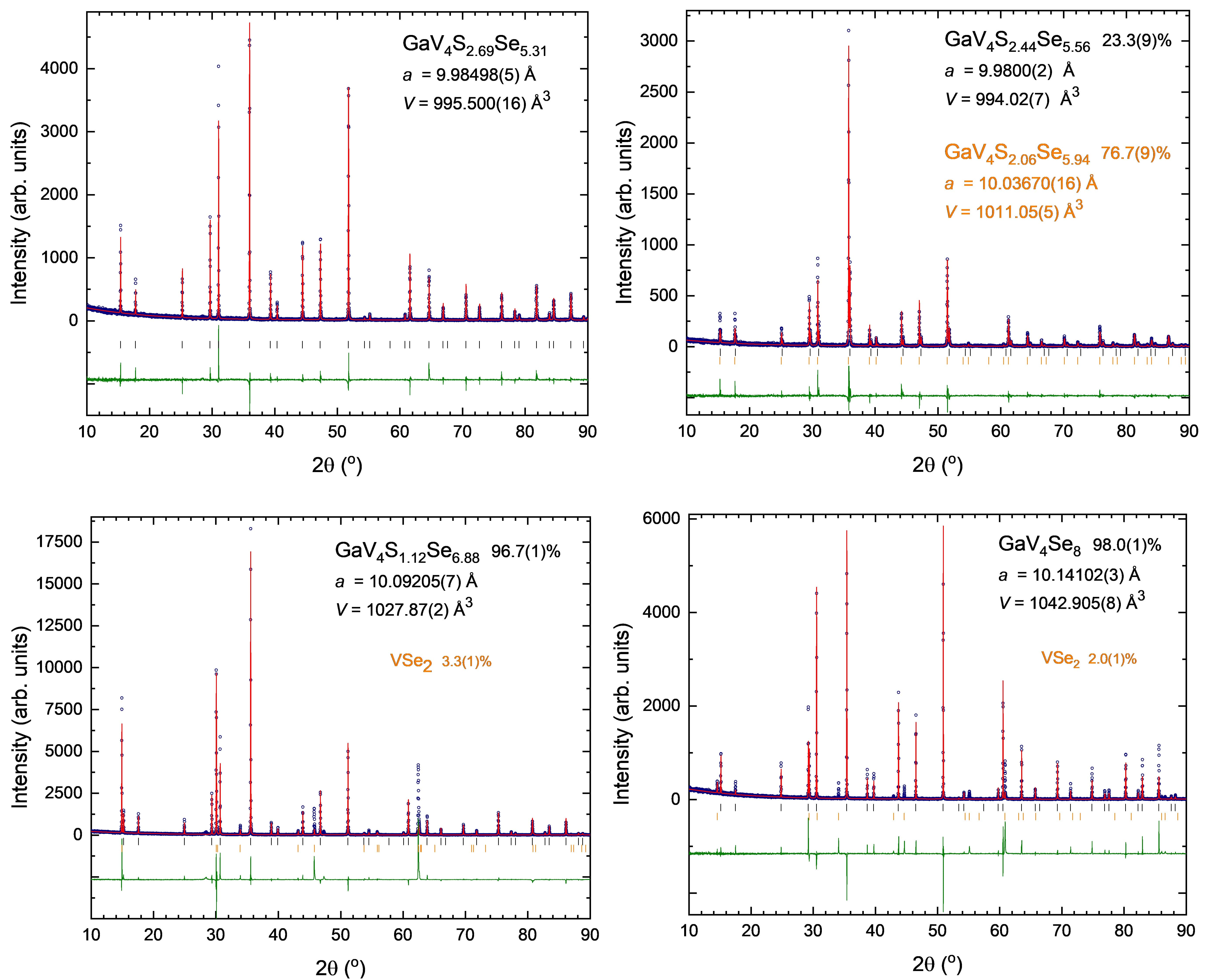}
\caption{Powder X-ray diffraction profiles of \ce{GaV4S3Se5}, \ce{GaV4S2S6}, \ce{GaV4SSe7}, and \ce{GaV4Se8}.
The experimentally-obtained diffraction profile at ambient temperature (blue open circles), refinement based on the model obtained from single crystal X-ray diffraction at room temperature (red solid line), difference (olive green solid line) and predicted peak positions (black tick marks).}
\end{figure}

\begin{figure}[t]
\includegraphics[width=0.75\linewidth]{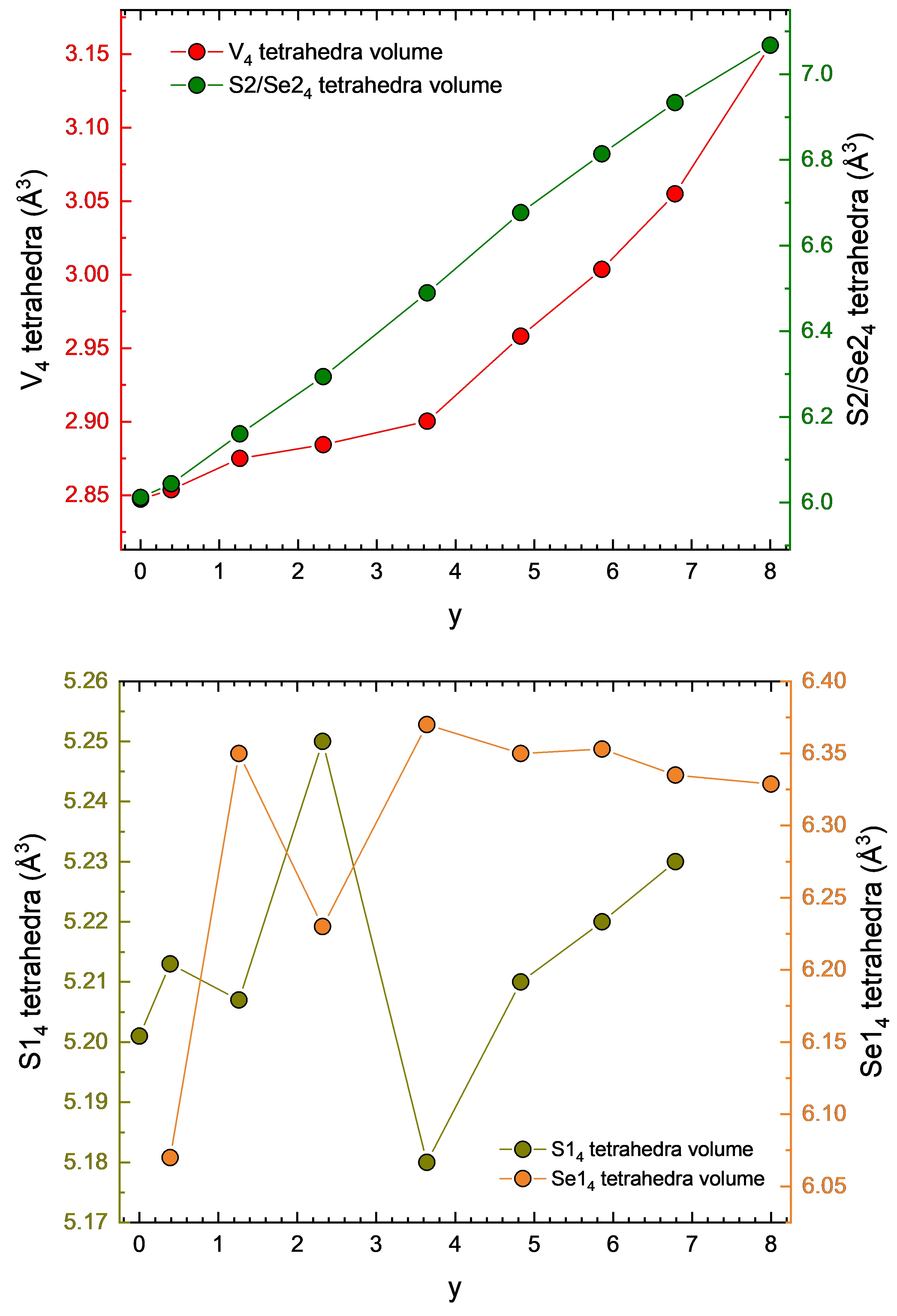}
\caption{Top: volume of the \ce{V4} tetrahedra and the S/Se site 2 tetrahedra. Bottom: volume of the S and Se tetrahedra within the V cluster for different compositions of GaV$_4$S$_{8-y}$Se$_y$.}
\end{figure}

\begin{figure}[t]
\includegraphics[width=\linewidth]{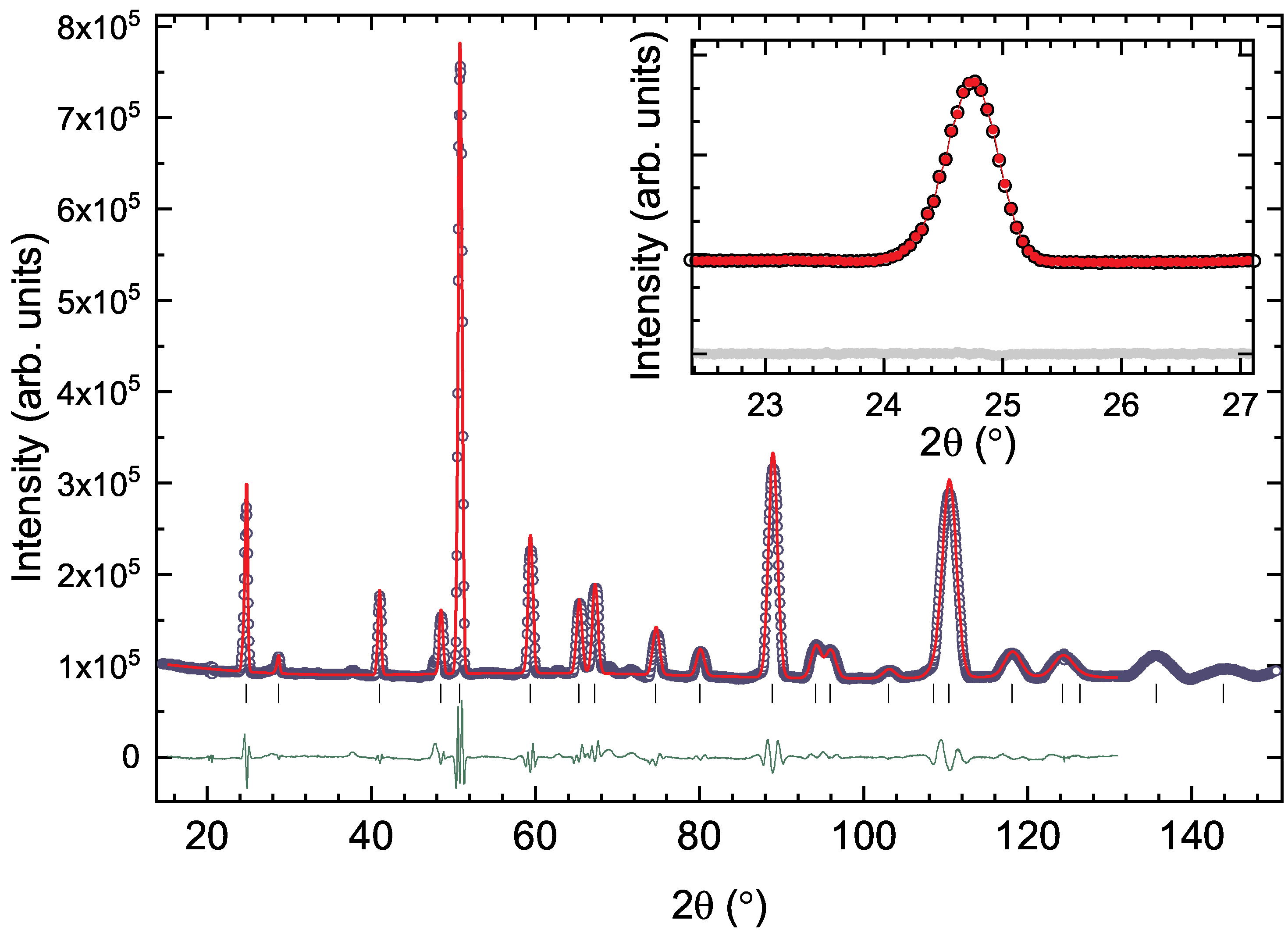}
\caption{\ce{GaV4S6Se2} powder neutron diffraction profile taken on the D20 diffratometer at the ILL.
The experimentally-obtained diffraction profile at 15~K (blue open circles), refinement based on the model obtained from single crystal X-ray diffraction at room temperature (red solid line), difference (olive green solid line) and predicted peak positions (black tick marks).
The inset shows the $(111)$ diffraction peak taken at 15~K (black open circles) and 1.5~K (red closed circles) and difference (grey closed circles).
At 1.5 and 15~K the lattice parameter refine in an $F\bar{4}3m$ to be $9.7673(4)$ and $9.7669(4)$ respectively.
}
\end{figure}

\begin{figure}[t]
\includegraphics[width=\linewidth]{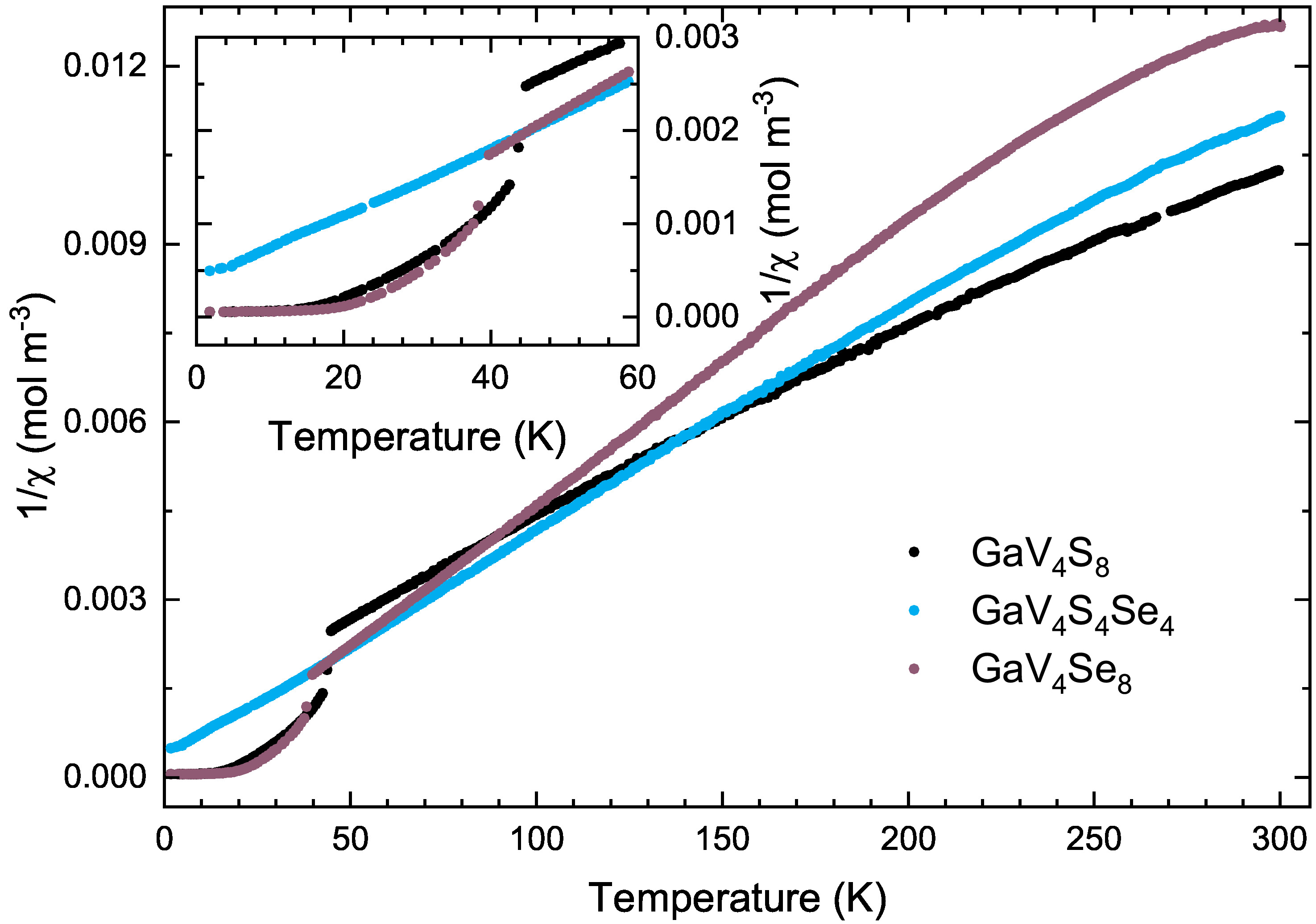}
\caption{Inverse magnetic susceptibility vs temperature for \ce{GaV4S8}, \ce{GaV4S4Se4}, and \ce{GaV4Se8} measured in an applied magnetic field of 10 mT.}
\end{figure}

\end{document}